\documentclass[letterpaper,aps,sort,amsfonts,amssymb,amsmath,prb,floatfix,
preprintnumbers,twocolumn,showpacs,showkeys]{revtex4}
\usepackage[T1]{fontenc}
\usepackage{amsmath}
\usepackage{color}
\usepackage{graphics}
\usepackage{amssymb}
\usepackage{pifont}

\makeatletter

\providecommand{\LyX}{L\kern-.1667em\lower.25em\hbox{Y}\kern-.125emX\@}
\let\SF@@footnote\footnote
\def\footnote{\ifx\protect\@typeset@protect
    \expandafter\SF@@footnote
  \else
    \expandafter\SF@gobble@opt
  \fi
}
\expandafter\def\csname SF@gobble@opt \endcsname{\@ifnextchar[
  \SF@gobble@twobracket
  \@gobble
}
\edef\SF@gobble@opt{\noexpand\protect
  \expandafter\noexpand\csname SF@gobble@opt \endcsname}
\def\SF@gobble@twobracket[#1]#2{}

\usepackage{graphicx}
\usepackage{dcolumn}
\usepackage{bm}

\makeatother
\AtBeginDocument{
  
}

\begin{document}

\preprint{preprint to be submitted to Phys. Rev. B}

\title{Evidence for strong electron-phonon coupling in MgCNi\( _{3} \)}

\author{A. W\"alte}

\author{G. Fuchs}

\email{fuchs@ifw-dresden.de (G.Fuchs)}

\author{K.-H. M\"uller}

\author{A. Handstein }

\author{K. Nenkov}

\thanks{On leave from: Int Lab. of High Magn. Fields, Wroclaw; ISSP-BAS,
Sofia, Bulgaria.}

\author{V.N. Narozhnyi}

\thanks{On leave from Inst. for High Pressure Physics, Troitsk, 142190, 
Russia.}

\author{S.-L. Drechsler}

\author{S. Shulga}

\thanks{On leave from Inst. of Spectroscopy, Troitsk, 142190, Russia.}

\author{L. Schultz}

\affiliation{Institut für Festk\"orper- und Werkstoffforschung Dresden, 
Postfach
270116, D-01171 Dresden, Germany}

\author{H. Rosner}

\affiliation{Max-Planck-Institut für Chemische Physik fester Stoffe, 
N\"othnitzer
Strasse 40, D-01187 Dresden, Germany}

\date{1 August 2002}

\begin{abstract}
The title compound is investigated by specific heat measurements in
the normal and superconducting state supplemented by upper critical
field transport, susceptibility and magnetization measurements. From
a detailed analysis including also full potential electronic structure
calculations for the Fermi surface sheets, Fermi velocities and partial
densities of states the presence of both strong electron-phonon interactions
and considerable pair-breaking has been revealed. The specific heat
and the upper critical field data can be described to first approximation
by an effective single band model close to the clean limit derived
from a strongly coupled predominant hole subsystem with small Fermi
velocities. However, in order to account also for Hall-conductivity
and thermopower data in the literature, an effective general two-band
model is proposed. This two-band model provides a flexible enough
frame to describe consistently all available data within a scenario
of phonon mediated \( s \)-wave superconductivity somewhat suppressed
by sizeable electron-paramagnon or electron-electron Coulomb interaction.
For quantitative details the relevance of soft phonons and of a van
Hove type singularity in the electronic density of states near the
Fermi energy is suggested.
\end{abstract}



\maketitle

\section{Introduction\label{Sec - Introduction}}

\renewcommand{\thefootnote}{\roman{footnote}}{The
recent discovery of superconductivity in the intermetallic antiperovskite
compound MgCNi\( _{3} \)\cite{he01} with a superconducting transition
temperature of \( T_{\textrm{c}}\simeq 8\textrm{ K} \) is rather
surprising considering its high Ni content. Therefore it is expected
that this compound is near a ferromagnetic instability which might
be reached by hole doping on the \( \textrm{Mg} \) sites.\cite{rosner02}
The possibility of unconventional superconductivity due to the proximity
of these two types of collective order has attracted great interest
in the electronic structure and the physics of the pairing mechanism.}

Band structure calculations\cite{dugdale01,singh01,Szajek01,rosner02,shim01}
for MgCNi\( _{3} \) revealed a domination of the electronic states
at the Fermi surface by the \( 3d \) orbitals of \( \textrm{Ni} \),
suggesting presence of ferromagnetic spin fluctuations.\cite{singh01,rosner02}
\( ^{13}\textrm{C NMR} \) measurements\cite{singer01} \textcolor{black}{result
in Fer}mi liquid behavior with an electronic crossover at 
\( T\approx 50\textrm{ K} \),
growing formation of spin fluctuations below \( T\approx 20\textrm{ K} \).
Resistivity measurements,\cite{li01,he01,kumary02} measurements of
the thermopower, the thermal conductivity and the magnetoresistance,\cite{li02}
doping experiments\cite{hayward01,kumary02} and magnetization 
measurements\cite{hayward01} are consistent with this interpretation. 

\( \textrm{MgCNi}_{3} \) can be considered as the \( 3 \)-dimensional
analogue of the quaternary layered transition metal borocarbides which
exhibit superconducting transition temperatures up to 
\( T_{\textrm{c}}\simeq 23\textrm{ K} \).
In spite of the much lower \( T_{\textrm{c}} \) of \( \textrm{MgCNi}_{3} \),
its upper critical field \( H_{\textrm{c}2} \) at low 
temper\textcolor{black}{atures,
\( H_{\textrm{c}2}(0)=8\ldots 15\textrm{ T} \),
\cite{li01,li01a,mao03,young03,lin03}
is comparab}le with that of the borocarbides or even higher. However,
a rather different shape, especially near \( T_{\textrm{c}} \), for
the temperature dependence of \( H_{\textrm{c}2}(T) \) is observed
for these compounds. The \( H_{\textrm{c}2}(T) \) dependence of 
\( \textrm{MgCNi}_{3} \)
is similar to that of usual (standard) intermetallic superconductors
which are described reasonably well within the isotropic single-band
approximation and exhibit a steep slope of \( H_{\textrm{c}2}(T) \)
at \( T_{\textrm{c}} \).

Through analysis of specific heat data, \( \textrm{MgCNi}_{3} \)
was characterized in the framework of a conventional, phonon-mediated
pairing both as a moderate \cite{he01,lin03} and as a 
strong\cite{shan03,mao03}
coupling superconductor. Strong coupling is also suggested by measurements
of the thermopower\cite{li02} and the large energy gap determined
from tunneling experiments.\cite{mao03} The question of the pairing
symmetry is controversially discussed in the literature. \( ^{13} \)C
NMR experiments,\cite{singer01} specific heat measurements\cite{lin03}
and tunneling spectra\cite{shan03pc} support \( s \)-wave pairing
in \( \textrm{MgCNi}_{3} \), whereas earlier tunneling spectra\cite{mao03}
and penetration depth measurements\cite{prozorov03} have been interpreted
in terms of an unconventional pairing sta\textcolor{black}{te. Recent
measurements of the critical current of \( \textrm{MgCNi}_{3} \)
may be interpreted in the latter sense, too.\cite{Young_cond-mat03}}

In the present investigation, specific heat data of \( \textrm{MgCNi}_{3} \)
in the normal and superconducting state were analyzed in detail with
the aid of a realistic phonon model and strong coupling corrections
as suggested by \citeauthor{carbotte90}.\cite{carbotte90} The results
are brought into accordance with the two-band character of 
\( \textrm{MgCNi}_{3} \)
emerging from band structure calculations and a parallel analysis
of the upper critical field \( H_{c2}(0) \), in order to find out
a consistent physical picture explaining at least qualitatively various
available experimental results.

\section{\label{Sec 2}Essentials of the theoretical electronic structure}

\begin{figure}
{\centering \resizebox*{0.48\textwidth}{!}{\includegraphics{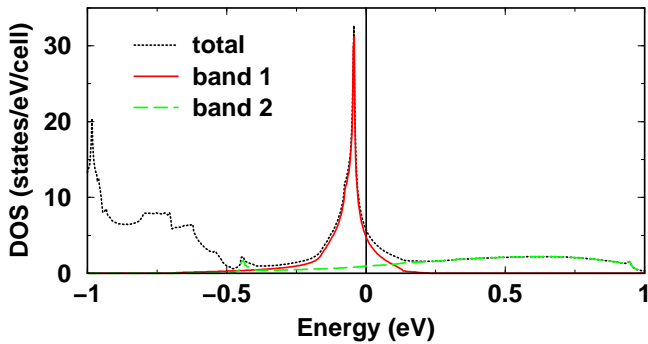}} \par}

\caption{Partial density of states of the two bands in 
\protect\( \textrm{MgCNi}_{3}\protect \)
corresponding to the two Fermi surface sheets shown in 
Fig. \ref{Bild - surface sheets 1}.
Dotted line: total density of states.\label{Bild - DOS}}
\end{figure}
 
\begin{figure}
{\centering \resizebox*{0.48\textwidth}{!}{\includegraphics{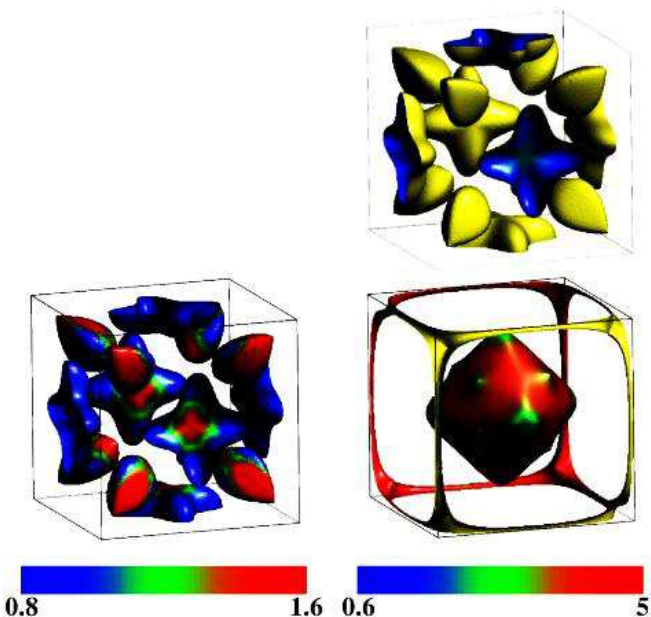}} \par}

\caption{\textcolor{black}{Fermi surface sheets of 
\protect\( \textrm{MgCNi}_{3}\protect \).
Fermi velocities are measured in different colors (see scales below
the figure) in units of \protect\( 10^{7}\textrm{ cm}/\textrm{s}\protect \),
i.e. blue color stands for slow and red color for fast quasiparticles.
Upper panel and lower left panel: hole sheets corresponding to {}``band
1'' in Fig. \ref{Bild - DOS}. Lower right panel: electron sheets
corresponding to {}``band 2'' in Fig. \ref{Bild - DOS}. The right
panels present the Fermi velocity distribution of the two sheets on
the same absolute scale to demonstrate the slow (heavy) character
of the holes. Yellow color: sides of filled electrons. The left panel
shows the \protect\( v_{\textrm{F}}\protect \)-distribution in the
hole sheets on a smaller scale in more detail.}
\textcolor{red}{\label{Bild - surface sheets 1} }}
\end{figure}
 Following previous work of one of the authors (H.R.)\cite{rosner02}
in the present section we remind the reader on some essential 
feat\textcolor{black}{ures
and point out i}mportant, but nevertheless so far unpublished details
of the electronic structure of \( \textrm{MgCNi}_{3} \) which are
crucial for a proper interpretation of the specific heat (total and
Fermi surface sheet (FSS) related partial densities of states (DOS)),
upper critical field and transport data (topology of the Fermi surface
and the magnitude of the Fermi velocities). Among various band structure
calculations there is general consensus about the qualitative topology
of the Fermi surface and the presence of a strong peak (Van Hove Singularity)
in \( N(E) \) slightly below the Fermi energy. At the same time there
are clear differences with respect to the magnitudes of 
\( N(0)=4.8\textrm{ states}/\textrm{eV}=11\textrm{ mJ}/\textrm{molK}^{2} \)
(to be compared with \( 4.63\textrm{ states}/\textrm{eV} \),\cite{dugdale01}
\( 4.99\textrm{ states}/\textrm{eV} \),\cite{singh01} 
\( 5.34\textrm{ states}/\textrm{eV} \)\cite{shim01})
and especially with respect to the Stoner factor \( \textrm{S}=3.3 \)
(compared with \( 1.75 \),\cite{dugdale01} \( 2.78 \),\cite{shein02}
to \( 5 \),\cite{singh01}) as well as to the distance of the DOS
peak \( 42\textrm{ meV} \) (compared with \( 40\textrm{ meV} \)
\cite{Ignatov03}
to \( 80\textrm{ meV} \)\cite{shein02}) below \( E_{\textrm{F}} \).
The peak may be of relevance for a proper quantitative description
of the electronic specific heat, transport data, magnetic properties,
and the superconductivity. Last but not least, there is also a sizeable
disparity on the magnitude of the electron-phonon coupling constant
\( \lambda _{\textrm{ph}} \) (ranging between \( 0.8\ldots 2.0 \))
mainly caused by the poor knowledge of the phonon energies and possible
lattice anharmonicities.\cite{Ignatov03}

Our results have been obtained by a band structure calculation code
using the full-potential nonorthogonal local-orbital (FPLO) minimum-basis
scheme.\cite{koepernik99} There are about \( 0.285 \) charges per
unit cell with exactly equal numbers of holes and electrons, i.e.
\( n_{\textrm{h}}=n_{\textrm{el}} \) which follows from the even
number of electrons per unit cell. In other words, \( \textrm{MgCNi}_{3} \)
is a so called compensated metal which must be described in terms
of multi-band model by definition (per s$\acute{\textrm{e}}$). Thus,
it makes sense to start with a two-band model. The generalization
to any higher multi-band scenario is straightforward. A standard single
band system with an even number of electrons per unit cell would be
an insulator and not a superconductor. Thus metalicity is achieved
owing to the two-band character which leads to electron and hole derived
Fermi surface sheets (FSS). The total DOS \( N(0) \) at the Fermi
level can be decomposed into a roughly \( 85\% \) and a \( 15\% \)
contribution stemming from two hole and two electron sheets of the
Fermi surface, respectively (see Figs. \ref{Bild - DOS}
, \ref{Bild - surface sheets 1}).
On the Fermi surface sheets shown in Fig. \ref{Bild - surface sheets 1}
the sides of filled electron states are shown in yellow/gold color.
The two types of hole sheets are formed by eight droplets (ovoids)
oriented along the spatial diagonals of the cube, i.e. along 
the \( \Gamma  \)-\( \textrm{R} \)
lines and six FSS with a {}``four-leaved clover''-like shape centered
at the X-points in the middle of the faces of the cube (see 
Fig. \ref{Bild - surface sheets 1}).
The coordinates of the symmetry points read \( \Gamma =(0,0,0) \),
\( \textrm{R}=(0.5,0.5,0.5) \), \( \textrm{X}=(0.5,0,0) \) and 
\( \textrm{M}=(0.5,0.5,0) \)
(all given in units of \( 2\pi /a \), where \( a=0.381\textrm{ nm} \)
is the lattice constant). The FSS with electron character are given
by the rounded cube centered at \( \Gamma  \) and \( 12 \) thin
jungle gims spanning from \( \textrm{R} \) to \( \textrm{M} \). 

The band structure calculations provide us directly with several material
parameters (total and partial densities of states, Fermi velocities,
etc.) important for the understanding of superconductivity and electronic
transport properties. For instance the transport properties are described
by quadratically averaged Fermi 
velocities: \( \left\langle v^{2}\right\rangle _{\textrm{FSS}} \)
whereas the upper critical field is described by averages of the type
\( \left\langle 1/v^{2}\right\rangle _{\textrm{FSS}} \) which yields
a smaller effective velocity in general. Using the general definitions
of the local density of states (in \( \overrightarrow{k} \)-space)
and those of \( \textrm{m}^{th} \) and the first moments of the Fermi
velocity \( v=\left| \overrightarrow{v}\left( \overrightarrow{k}\right) 
\right|  \),
respectively, we have\begin{eqnarray}
\left\langle v^{m}\right\rangle _{i} & = & 
\frac{\int \textrm{d}S_{i}N_{i}\left( \overrightarrow{k}\right) \left| 
v_{i}\left( \overrightarrow{k}\right) \right| ^{m-1}}{\int 
\textrm{d}S_{i}N_{i}\left( \overrightarrow{k}\right) }\nonumber \\
 & \qquad \equiv  & \frac{\int \textrm{d}S_{i}\left| \overrightarrow{v}_{i}
\left( \overrightarrow{k}\right) \right| ^{m-1}}{4\pi ^{3}\hbar N_{i}(0)},
\nonumber \\
\left\langle v\right\rangle _{i} & \equiv  & \overline{v}_{i}\nonumber \\
 & \qquad = & \frac{S_{F,i}}{4\pi ^{3}\hbar N_{i}(0)},\nonumber \\
\left\langle v^{m}\right\rangle _{i} & = & \frac{\overline{v}_{i}\int 
\textrm{d}S_{i}\left| \overrightarrow{v}\left( \overrightarrow{k}\right) 
\right| ^{m-1}}{S_{F,i}},\nonumber 
\end{eqnarray}
where \( i=el,h,tot \) and \( S_{f,i} \) denotes the area of the
\( i^{th} \) Fermi surface sheet and the effective quantity is related
to the linearly averaged value \( \overline{v} \) adopting a simple
estimate as\begin{eqnarray}
\left\langle v^{-2}\right\rangle _{i} & = & \frac{\overline{v}_{i}\int 
\textrm{d}S_{i}\left| \overrightarrow{v}_{i}\left( \overrightarrow{k}\right) 
\right| ^{-3}}{S_{F,i}}\equiv v_{\textrm{hc}2,i}^{-2},\nonumber \\
v_{\textrm{hc}2} & \approx  & \overline{v}\left[ 1-\left( \delta v/
\overline{v}\right) ^{2}\right] ,\nonumber 
\end{eqnarray}
where \( \delta v \) is the halfwidth of the \( v_{\textrm{F}} \)
distribution. For the two above mentioned subgroups of quasiparticles
we estimate: \( \overline{v}_{2}=3.9\times 10^{7}\textrm{ cm}/\textrm{s} \)
and \( \overline{v}_{1}=1.2\times 10^{7}\textrm{ cm}/\textrm{s} \),
\( v_{\textrm{tr},1}=1.11\times 10^{7}\textrm{ cm}/\textrm{s} \),
\( v_{\textrm{hc}2,1}=1.07\times 10^{7}\textrm{ cm}/\textrm{s} \)
and \( v_{\textrm{tr},2}=4.89\times 10^{7}\textrm{ cm}/\textrm{s} \),
where \( \delta v_{1}=4\times 10^{6}\textrm{ cm}/\textrm{s} \) and
\( \delta v_{2}=1.1\times 10^{7}\textrm{ cm}/\textrm{s} \) have been
used (compare also Fig. \ref{Bild - surface sheets 1}). 

Finally, in the isotropic single band (ISB) model realized in the
extreme dirty limit of superconductivity one arrives at\[
N(0)v^{2}_{\textrm{tr},\textrm{ISB}}=N_{1}(0)v^{2}_{
\textrm{tr},1}+N_{2}(0)v^{2}_{\textrm{tr},2},\]
which yields \( v_{\textrm{tr},\textrm{ISB}}=2.15\times 10^{7}\textrm{ cm}/
\textrm{s} \)
in accordance with Ref. \onlinecite{dugdale01}. The corresponding
\textcolor{black}{plasma energy amo}unts \( \hbar \omega _{\textrm{pl}}=3.17
\textrm{ eV} \)
close to \( 3.25\textrm{ eV} \) given in Ref. \onlinecite{singh01}.
Naturally, the total plasma frequency \( \omega _{\textrm{pl}} \)
can be also decomposed into the plasma frequencies of both subsystems\[
\omega ^{2}_{\textrm{pl}}=\omega ^{2}_{\textrm{pl},1}+
\omega ^{2}_{\textrm{pl},2}=\omega ^{2}_{\textrm{pl},1}
\left( 1+\frac{N_{2}(0)v^{2}_{\textrm{tr},2}}{N_{1}(0)
v^{2}_{\textrm{tr},1}}\right) .\]
Thus we estimate \( \hbar \omega _{\textrm{pl},1}\equiv \hbar 
\omega _{\textrm{pl},\textrm{h}}\approx 1.89\ldots 1.94\textrm{ eV} \)
and \( \hbar \omega _{\textrm{pl},2}\equiv \hbar 
\omega _{\textrm{pl},\textrm{el}}\approx 2.55\ldots 2.61\textrm{ eV} \).
From these partial p\textcolor{black}{lasma energies a useful re}lation
between the scattering rates \( \gamma _{\textrm{i}} \) and the conductivities
\( \sigma _{\textrm{i}} \) (with \( \textrm{i}=1,2 \)) in both subsystems
can be obtained:

\[
\frac{\gamma _{\textrm{el}}}{\gamma _{\textrm{h}}}=1.816
\frac{\sigma _{\textrm{h}}}{\sigma _{\textrm{el}}}=1.816
\frac{\rho _{\textrm{el}}}{\rho _{\textrm{h}}},\]
where \( \rho _{\textrm{i}} \) denotes the corresponding resistivity.
In the present case the disorder is expected to be caused mainly by
Mg and C related defects such as vacancies and interstitials. Therefore
the ratio of the scattering rates might scale with the ratio of the
non-Ni derived Mg and C orbital partial densities of states at the
Fermi level and the corresponding Fermi velocities. As a result we
estimate from our LDA-FPLO calculations

\begin{equation}
\label{Eq - scatteringrateratio}
\left( \frac{\gamma _{\textrm{el}}}{\gamma _{\textrm{h}}}\right) _{
\textrm{LDA}}\approx \frac{N_{\textrm{el},\textrm{Mg},\textrm{C}}(0)
v_{\textrm{F},\textrm{el}}}{N_{\textrm{h},\textrm{Mg},\textrm{C}}(0)
v_{\textrm{F},\textrm{h}}}\approx 4.81.
\end{equation}
Within this approach the corresponding mean free paths differ by a
factor of \( 0.917 \) and a conductivit\textcolor{black}{y ratio
of \( \sigma _{\textrm{h}}/\sigma _{\textrm{el}}=1.403 \) would be
expected. }

\textcolor{black}{In the following analysis we usually make use of
\( \hbar =k_{\textrm{B}}=\mu _{0}=1 \) for the sake of simplification.}

\section{{\normalsize Experimental\label{Sec - experimental}}}

\begin{figure}
{\centering \resizebox*{0.48\textwidth}{!}{\includegraphics{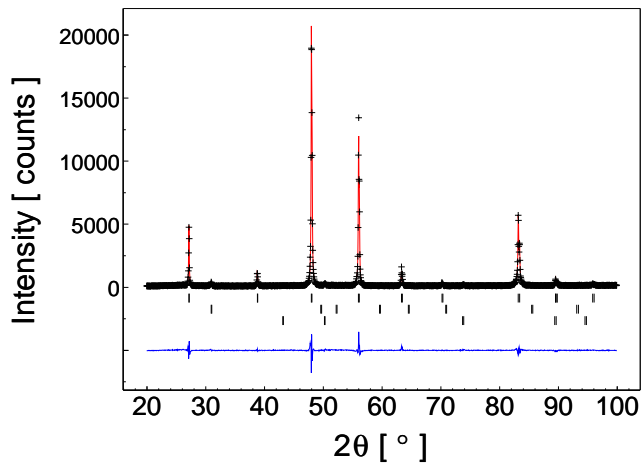}} \par}

\caption{Rietveld refinement for the \protect\( \textrm{MgC}_{1.6}
\textrm{Ni}_{3}\protect \)
sample. The crosses correspond to the experimental data. The black
line shows the calculated pattern. The vertical bars give the Bragg
positions for the main phase \protect\( \textrm{MgCNi}_{3}\protect \),
for graphite and \protect\( \textrm{MgO}\protect \) (from top to
bottom). The black line at the bottom of the plot gives the difference
between the experimental and calculated pattern. \label{xray Diffraktogramm}}
\end{figure}

Polycrystalline samples of MgCNi\( _{3} \) have been prepared by
solid state reaction. In order to obtain samples with high 
\( T_{\textrm{c}} \),
we used an excess of carbon \textcolor{black}{as proposed in} Ref.
\onlinecite{he01}. To cover the high volatility of Mg during sintering
of the samples an excess of Mg is used.\cite{he01} In this study,
a sample with the nominal formula \( \textrm{Mg}_{1.2}
\textrm{C}_{1.6}\textrm{Ni}_{3} \)
has been investigated and is denoted as \( \textrm{MgC}_{1.6}
\textrm{Ni}_{3} \).
To prepare the sample a mixture of Mg, C and Ni powders was pressed
into a pellet. The pellet was wrapped in a Ta foil and sealed in a
quartz ampoule containing an Ar atmosphere at \( 180\textrm{ mbar} \).
The sample was sintered for half an hour at \( 600^{\circ }\textrm{C} \)
followed by one hour at \( 900^{\circ }\textrm{C} \). After a cooling
process the sample was reground. This procedure was repeated two times
in order to lower a possible impurity phase content. The obtained
sample was investigated by x-ray diffractometry to estimate its quality.
The diffractometer pattern (Fig. \ref{xray Diffraktogramm}) shows
small impurity concentrations mainly resulting from MgO and  unreacted
carbon crystallized as graphite (\( \approx 10 \) vol.-\%). The lattice
constant of the prepared sample was determined to be 
\( a=0.38107(1)\textrm{ nm} \)
using the Rietveld code {\small FULLPROF}.\cite{fullprof02} This
indicates that the nearly single-phase sample corresponds to the 
superconducting
modification of \( \textrm{MgC}_{\textrm{x}}\textrm{Ni}_{3} \).\cite{ren02}
The superconducting transition of the sample was investigated by measurements
of electrical resistance, ac susceptibility and specific heat. For
the electrical resistance measurement a piece cut from the initially
prepared pellet with \( 5\textrm{ mm} \) in length and a cross section
of approximately \( 1\textrm{ mm}^{2} \) was measured in magnetic
fields up to \( 16\textrm{ T} \) using the standard four probe method
with current densities between \( 0.2\textrm{ and }1\textrm{ A}/
\textrm{cm}^{2} \).
The ac susceptibility and the specific heat measurements were performed
on other pieces from the same pellet in magnetic fields up to 
\( 9\textrm{ T} \).

\section{Results}

\subsection{Resistivity\label{Section 4.1}}

\begin{figure}
{\centering \resizebox*{0.49\textwidth}{!}{\includegraphics{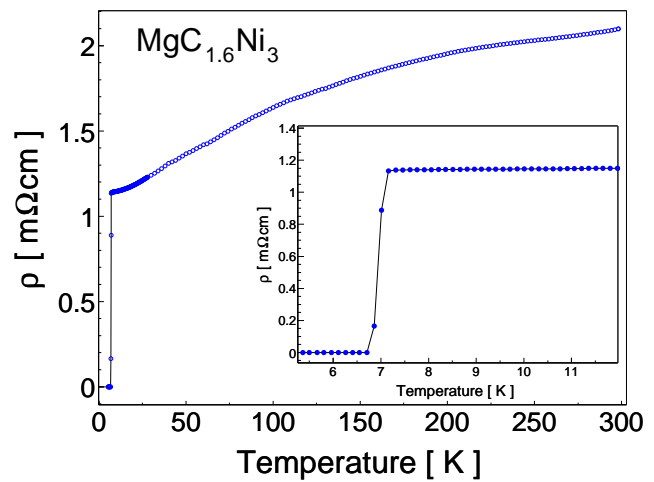}} \par}

\caption{Resistivity as a function of temperature of the \protect\( 
\textrm{MgC}_{1.6}\textrm{Ni}_{3}\protect \)
sample up to room temperature. The inset shows the superconducting
transition region.\label{Bild - Gesamtwiderstand}}
\end{figure}

The temperature dependence of the electrical resistance of the investigated
sample is shown in Fig. \ref{Bild - Gesamtwiderstand}. A superconducting
transition with an onset (midpoint) value of \( T_{\textrm{c}}=7.0
\textrm{ K }\left( 6.9\textrm{ K}\right)  \)
is observed (see inset of Fig. \ref{Bild - Gesamtwiderstand}) which
coincides with the onset of the superconducting transition of \( T_{
\textrm{c}}=7.0\textrm{ K} \)
determined from ac susceptibility. Its residual resistance ratio \( 
\rho (300\textrm{K})/\rho (8\textrm{K})=1.85 \)
and the shape of the \( \rho (T) \) curve are typical for MgCNi\( _{3} \)
powder samples.\cite{he01,kumary02,li01} It should be noted, that
the sample of Fig. \ref{Bild - Gesamtwiderstand} has a resistivity
of \( \rho _{300\textrm{K}}=2.1\textrm{ m}\Omega \textrm{cm} \) which
is much too large in order to be intrinsic.

A natural explanation for the high resistivity of the investigated
sample which was not subjected to high pressure sintering is a relatively
large resistance of the grain boundaries.\cite{kumary02} In this
context the recent low-resistivity thin film data (with \( \rho _{0} \)
down to \( 20\textrm{ }\mu \Omega \textrm{cm} \)) by \citeauthor{young03}
\cite{young03}
are of interest, since in that case similar values of the upper critical
field and \( T_{\textrm{c}} \) just as in the best powder samples\cite{li01}
have been observed.

\subsection{Specific heat\label{Sec - cp results} }

\begin{figure}
{\centering \resizebox*{0.48\textwidth}{!}{\includegraphics{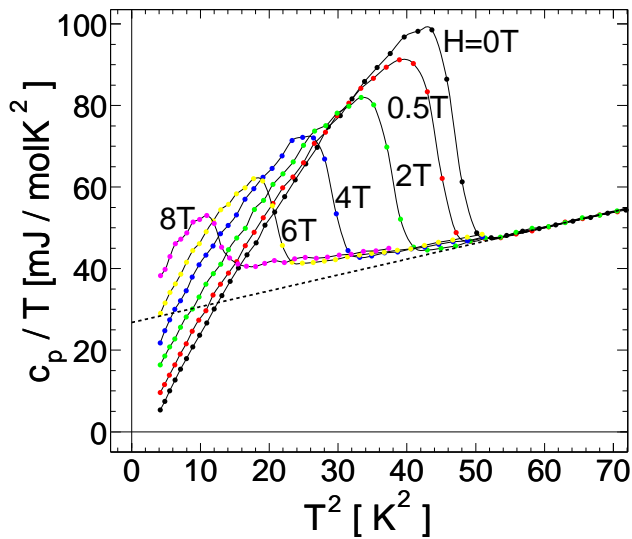}} \par}

\caption{Specific heat data \protect\( \textrm{c}_{\textrm{p}}/T\protect \)
vs. \protect\( T^{2}\protect \) of \protect\( \textrm{MgC}_{1.6}
\textrm{Ni}_{3}\protect \)
measured at various magnetic fields up to \protect\( 8\textrm{ T}\protect \).
The dashed line is a fit of the Debye approximation to the data for
\protect\( H=0\textrm{ T}\protect \) above \protect\( T_{\textrm{c}}
\protect \).
Its intersection with the \protect\( \textrm{c}_{\textrm{p}}/T\protect \)-axis
gives the Sommerfeld parameter \protect\( \gamma ^{\star }_{\textrm{N}}=27
\textrm{ mJ}/\textrm{molK}^{2}\protect \)
(see text).\label{Bild - spezifische W=E4rme komplett}}
\end{figure}

Specific heat \textcolor{black}{measurements were performed in order
to get information about the superconducting transition, the upper
critical field and the superconducting pairing symmetry and the strength
of the electron-phonon coupling from the}rmodynamic data. In Fig.
\ref{Bild - spezifische W=E4rme komplett} specific heat data, 
\( \textrm{c}_{\textrm{p}}/T \)
vs. \( T^{2} \), are shown for applied magnetic fields up to 
\( 8\textrm{ T} \).
The previously mentioned (see Sec. \ref{Sec - experimental}) \( 10 \)
vol.-\% graphite impurity contribution was subtracted according to
Ref. \onlinecite{Keesom55}.

The specific heat can be considered as a sum of a lattice contribution
and a linear-in-\( T \) term which gives the electronic contribution
with \( \gamma ^{\star }_{\textrm{N}} \) as the Sommerfeld parameter:
\begin{equation}
\label{spezifische Waerme cn}
\textrm{c}_{\textrm{n}}(T)=\gamma ^{\star }_{\textrm{N}}T+
\textrm{c}_{\textrm{lattice}}(T).
\end{equation}
To extract the lattice contribution of the normal state specific heat
the low temperature limit\begin{equation}
\label{Formel - Debye LT limit}
\textrm{c}_{\textrm{lattice}}(T)=\beta T^{3}
\end{equation}
of the Debye model is usually applied. A fit of 
Eq. (\ref{spezifische Waerme cn})
to the data is shown in Fig. \ref{Bild - spezifische W=E4rme komplett}
resulting the parameters \( \beta =0.39\textrm{ mJK}^{2}/\textrm{mol} \)
and \( \gamma ^{\star }_{\textrm{N}}=27.0\textrm{ mJ}/\textrm{molK}^{2} \).
Notice, that the Sommerfeld parameter is connected to the electron-phonon
coupling strength by \( \gamma ^{\star }_{\textrm{N}}=\gamma _{0}
\left( 1+\lambda _{\textrm{ph}}\right)  \).
With \( \gamma _{0}=11\textrm{ mJ}/\textrm{molK}^{2} \) (Sec. \ref{Sec 2}),
one obtains \( \lambda _{\textrm{ph}}=1.45 \) in contradiction with
recently reported medium coupling results.\cite{he01,lin03} From
the lattice contribution the Debye temperature \( \Theta ^{\star }_{
\textrm{D}}=292\textrm{ K} \)
was derived. Both parameters (\( \Theta _{\textrm{D}}^{\star } \),
\( \gamma _{\textrm{N}}^{\star } \)) are consistent with what has
been reported so far.\cite{mao03,lin03} The fit describes the normal
state data above \( T_{\textrm{c}} \) but its extrapolation to 
\( T=0\textrm{ K} \)
obviously underestimates the high-field data (see Fig. 
\ref{Bild - spezifische W=E4rme komplett}).

The transition temperature, \( T_{\textrm{c}}=6.8\textrm{ K} \),
calculated from entropy conservation criterion, agrees well with the
transition temperatures \( T_{\textrm{c}}=6.9\textrm{ K} \) and \( T_{
\textrm{c}}=7.0\textrm{ K} \)
derived from resistance and from ac susceptibility data, respectively. 

The jump \( \Delta \textrm{c}(T=T_{\textrm{c}}) \) of the specific
heat is given by the difference between the experimental data, \( 
\textrm{c}_{\textrm{p}}(T) \)
and the normal state specific heat contribution \( \textrm{c}_{
\textrm{n}}(T) \)\textcolor{black}{.
Notice that the experimental value of the jump, \( \Delta \textrm{c}(T_{
\textrm{c}})/\left( \gamma _{\textrm{N}}T_{\textrm{c}}\right) =2.09 \)
(derived from an entropy conserving construction -- see Sec. 
\ref{Sec - sl state analysis})
is strongly enhanced} compared to the BCS value \( (1.43) \), indicating
strong electron-phonon coupling.

\subsection{Superconducting transition and upper critical field
\label{Sec - transition&hc2}}

\textcolor{black}{The field dependence of the electrical resistance
of our investigated sample is shown in Fig. \ref{Bild - Widerstand vs 
Magnetfeld}
for several temperatures between \( 1.9\ldots 6.0\textrm{ K} \).
A sharp transition is observed. It remains sharp down to low temperatures.
In Fig. \ref{Bild - Widerstand & Suszeptibilit=E4t}, the field values
\( H_{10} \), \( H_{50} \) and \( H_{90} \) defined at \( 10\textrm{ }\% \),
\( 50\textrm{ }\% \) and \( 90\textrm{ }\% \) of the normal state
resistance are plotted as function of temperature. Identical results
have been found from resistance-vs.-temperature transition curves
measured at different magnetic fields. Additionally, Fig. \ref{Bild - 
Widerstand & Suszeptibilit=E4t}
shows upper critical field data determined from ac susceptibility
measurements, \( H^{\textrm{sus}}_{\textrm{c}2} \), determined by
an onset criterion. }
\begin{figure}
{\centering \resizebox*{0.49\textwidth}{!}{\includegraphics{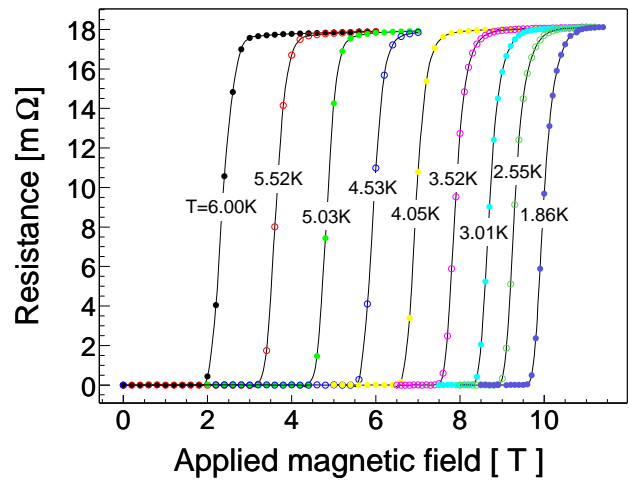}} \par}

\caption{\textcolor{black}{Resistivity of the \protect\( \textrm{MgC}_{1.6}
\textrm{Ni}_{3}\protect \)
sample as a function of the applied magnetic field for various fixed
temperatures as labeled.\label{Bild - Widerstand vs Magnetfeld}}}
\end{figure}

\textcolor{black}{It is clearly seen that for the investigated sample
\( H_{\textrm{c}2}^{\textrm{sus}} \) agrees approximately with \( H_{10} \).
A similar behavior was already observed for MgB\( _{2} \), whereas
in the case of rare-earth nickel borocarbides the onset of superconductivity
determined from ac susceptibility was typically found to agree well
with the midpoint value (\( H_{50} \)) of the normal state resistivity.
The width \( \Delta H=H_{90}-H_{10} \) of} the superconducting transition
curves in Fig. \ref{Bild - Widerstand vs Magnetfeld} (and Fig. \ref{Bild - 
Widerstand & Suszeptibilit=E4t})
remains, with \( \Delta H\simeq 0.6\textrm{ T} \), almost unchanged
down to low tempe\textcolor{black}{ratures. A non-textured polycrys}talline
sample of a strongly anisotropic superconductor shows a gradual broadening
of the superconducting transition with decreasing temperature as was
observed, for example, for \( \textrm{MgB}_{2} \).\cite{fuchs01}
Therefore, the nearly constant transition width \( \Delta H \) observed
for the investigated sample can be considered as an indication of
a rather small anisotropy of \( H_{\textrm{c}2}(T) \) in \( 
\textrm{MgCNi}_{3} \). 
\begin{figure}
{\centering \resizebox*{0.49\textwidth}{!}{\includegraphics{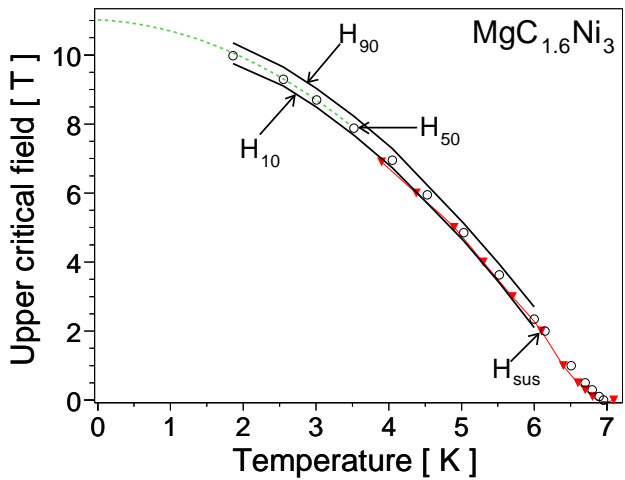}} \par}

\caption{The upper critical field as a function of temperature. The circles
show the midpoint of the transition (\protect\( \textrm{H}_{50}\protect \)).
The two lines labeled \protect\( \textrm{H}_{10}\protect \) and \protect\( 
\textrm{H}_{90}\protect \)
denote \protect\( 10\textrm{ }\%\protect \) and \protect\( 90
\textrm{ }\%\protect \)
of normal state resistivity. The triangles represent the upper critical
field from susceptibility measurements (onset values). The dashed
line illustrates the extrapolation of the resistivity data to 
\protect\( T=0\textrm{ K}\protect \).
\label{Bild - Widerstand & Suszeptibilit=E4t}}
\end{figure}

\begin{figure}
{\centering \resizebox*{0.49\textwidth}{!}{\includegraphics{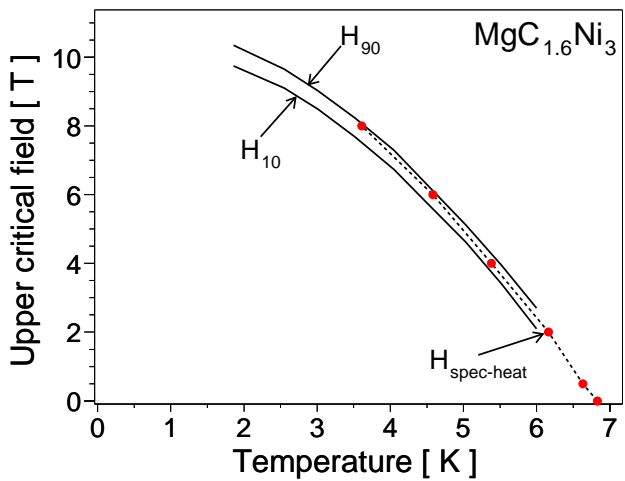}} \par}

\caption{Comparison of upper critical field data determined from specific
heat (\textcolor{red}{\tiny \ding{108}}) and resistance measurements.
\protect\( H_{10}\protect \) and \protect\( H_{90}\protect \) were
determined at \protect\( 10\%\protect \) and \protect\( 90\%\protect \)
of the normal state resistivity, respectively. An entropy conserving
construction was used to determine the upper critical field from the
specific heat data of Fig. \ref{Bild - spezifische W=E4rme komplett}.
\label{Bild - Widerstand & spezifische W=E4rme}}
\end{figure}

The upper critical fields, \( \textrm{H}_{\textrm{c}2}(T) \), determined
from the specific heat data, are shown in Fig. 
\ref{Bild - Widerstand & spezifische W=E4rme}.
The \( H_{c2}(T) \) data obtained from the specific heat are located
in the small field range between the \( H_{90}(T) \) and \( H_{10}(T) \)
curves determined from resistivity measurements (see Fig. 
\ref{Bild - Widerstand vs Magnetfeld}).

The extrapolation of \( H_{90}(T) \) to \( T=0\textrm{ K} \) yields
an upper critical field of \( H_{\textrm{c}2}(0)\simeq 11.0\textrm{ T} \)
(see Fig. \ref{Bild - Widerstand & Suszeptibilit=E4t}). The observed
temperature dependence of the upper critical field is typical for
\( H_{\textrm{c}2}(T) \) data reported for \( \textrm{MgCNi}_{3} \)
so far\cite{shan03} and was described\cite{li01,mao03,lin03} within
the standard WHH model\cite{werthammer66} by conventional superconductivity.
However, a quantitative analysi\textcolor{black}{s of \( H_{\textrm{c}2} \)
data pre}sented in Sec. \ref{Sec - Hc2 analysis} shows that the magnitude
of the upper critical field \( H_{\textrm{c}2}(0) \) at \( T=0\textrm{ K} \)
can be understood only if strong electron-phonon coupling is taken
into account.

\section{Analysis}

\subsection{Resistivity in the normal state}

The non-intrinsic origin of the residual resistivity \( \rho _{0\textrm{K}}
\approx 1.13\textrm{ m}\Omega \textrm{cm} \)
follows from physically reasonable values for the mean free path \( l \)
given by\[
l_{\textrm{imp}}=\frac{4\pi v_{\textrm{F}}}{\omega ^{2}_{\textrm{pl}}
\rho _{0}}\]
or in more convenient practical units\begin{equation}
\label{Formel - mittlere freie Weglaenge 2}
l_{\textrm{imp}}[\textrm{nm}]=4.9\times 10^{2}\frac{v_{\textrm{F}}
\left[ 10^{7}\textrm{ cm}/\textrm{s}\right] }{\left( \hbar \omega _{
\textrm{pl}}\left[ \textrm{eV}\right] \right) ^{2}\rho _{0}\left[ \mu 
\Omega \textrm{cm}\right] },
\end{equation}
with \( \rho _{0} \) \textcolor{black}{as the r}esidual resistivity.
In the isotropic single band (ISB) case, with \( v_{\textrm{F}}=v_{
\textrm{tr},\textrm{ISB}}\approx 2\times 10^{7}\textrm{ cm}/\textrm{s} \)
and the plasma energy \( \hbar \omega _{\textrm{pl}}\approx 3\textrm{ eV} \)
(see Sec. \ref{Sec 2}), one arrives at the conclusion, that in fact,
in the \( \textrm{m}\Omega \textrm{cm} \) range, typical for most
powder samples considered so far in the literature, even at 
\( T=0\textrm{ K} \)
the mean free path \( l \) -- given by 
Eq. (\ref{Formel - mittlere freie Weglaenge 2})
-- would be smaller than the lattice constant \( a
\approx 0.38\textrm{ nm} \)
in obvious conflict with the well-known Joffe-Regel limit \( l\geq a \).
\cite{ioffe60,mott72}
In other words the maximal intrinsic resistivity is given by \( \rho ^{
\textrm{max}}_{0}\left[ \textrm{m}\Omega \textrm{cm}\right] \approx 1.3v_{
\textrm{F}}\left[ 10^{7}\textrm{ cm}/\textrm{s}\right] /\left( \hbar 
\omega _{\textrm{pl}}\left[ \textrm{eV}\right] \right) ^{2}=0.29
\textrm{ m}\Omega \textrm{cm} \)
in the ISB dirty limit. Such resistivities \( \left( 0.33\textrm{ m}
\Omega \textrm{cm}\right)  \)
have been reported in most heavily (neutron) irradiated samples by
\citeauthor{karkin02}\cite{karkin02}

\subsection{Specific heat in the normal state
\label{Sec - normal state cp analysis}}

In order to describe the specific heat data in the normal state in
an extended temperature range \( T_{\textrm{c}}<T<30\textrm{ K} \),
the Debye low temperature limit approximation for the lattice contribution
(see Eq. (\ref{Formel - Debye LT limit})) was replaced by\[
\textrm{c}_{\textrm{lattice}}(T)=\textrm{c}_{\textrm{D}}(T)+
\textrm{c}_{\textrm{E}}(T).\]
Here,\[
\textrm{c}_{\textrm{D}}(T)=\Sigma ^{3}_{i=1}\; 3R\left( \frac{T}{\Theta _{
\textrm{D}i}}\right) ^{3}\int ^{\Theta _{\textrm{D}i}/T}_{0}\textrm{d}x\frac{
\textrm{e}^{x}x^{4}}{\left( \textrm{e}^{x}-1\right) ^{2}}\]
stands for the Debye model\cite{srivastava90,kittel} describing the
\( 3 \) acoustic phonon branches, whereas the Einstein model
\cite{srivastava90,kittel}\[
\textrm{c}_{\textrm{E}}(T)=\Sigma ^{15}_{i=4}\; R\left( \frac{\Theta _{
\textrm{E}i}}{T}\right) ^{2}\frac{\exp \left( \Theta _{\textrm{E}i}/T
\right) }{\left[ \exp \left( \Theta _{\textrm{E}i}/T\right) -1\right] ^{2}}\]
describes the \( 12 \) optical branches.

\begin{figure}
{\centering \resizebox*{0.48\textwidth}{!}{\includegraphics{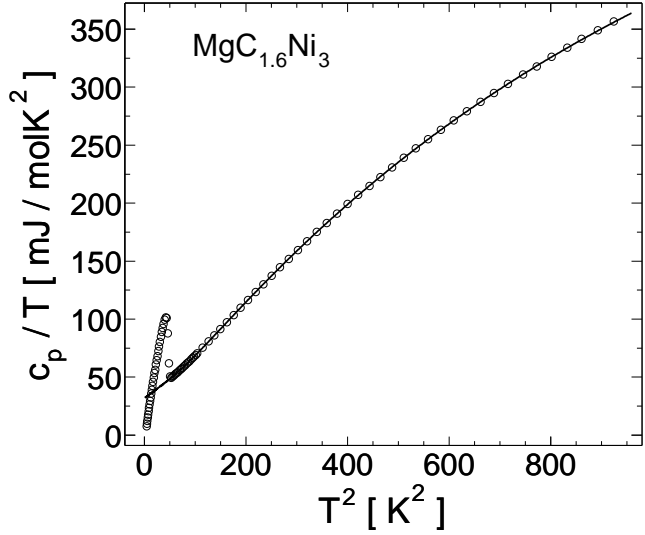}} \par}

\caption{Specific heat data \protect\( \textrm{c}_{\textrm{p}}(T)/T\protect \)
vs. \protect\( T^{2}\protect \) for zero magnetic field in the temperature
range up to \protect\( 30\textrm{ K}\protect \). The solid line is
a fit of the lattice model (see text for details), showing very good
agreement with the data for \protect\( T_{\textrm{c}}<T<30\textrm{ K}
\protect \).\label{Bild - spezifische W=E4rme H=3D0}}
\end{figure}

\begin{table}

\caption{\textcolor{black}{Debye and Einstein temperatures with corresponding
occupati}on numbers. \protect\( \textrm{D}i\protect \) denote the
acoustic phonons and \protect\( \textrm{E}i\protect \) the optical
phonons. \protect\( \Theta \protect \) gives the corresponding temperature
and \protect\( \nu _{\textrm{i}}\protect \) is the grouping parameter,
giving the number of modes found to have the same temperature.
\label{Tabelle - Gitterfitparameter}}

\begin{ruledtabular}

{\centering \begin{tabular}{cp{0.6cm}ddddddd}
&
&
\multicolumn{2}{c}{acoustic modes}&
\multicolumn{5}{c}{optical modes}\\
&
&
\textrm{D1}&
\textrm{D2}&
\textrm{E1}&
\textrm{E2}&
\textrm{E3}&
\textrm{E4}&
\multicolumn{1}{p{0.8cm}}{\textrm{E5}}\\
\hline 
\( \Theta  \)&
\multicolumn{1}{l}{\( \left[ \textrm{K}\right]  \)}&
129&
316&
86&
163&
256&
472&
661\\
\( \nu _{i} \)&
&
1&
2&
0.33&
2.67&
3&
3&
3\\
\end{tabular}\par}

\end{ruledtabular}
\end{table}
We found that the \( 9 \) energetically lowest phonons (\( 3 \)
acoustic and \( 6 \) optical modes) are sufficient to describe the
normal state specific heat up to \( T=30\textrm{ K} \). The result
of the fit is shown in Fig. \ref{Bild - spezifische W=E4rme H=3D0}.
The Sommerfeld parameter converged to \( \gamma ^{\star }_{\textrm{N}}=31.4
\textrm{ mJ}/\textrm{molK}^{2} \),
greater than determined from Fig. \ref{Bild - spezifische W=E4rme komplett}.
Specific heat measurements up to \( T=300\textrm{ K} \) on another
piece from the initially prepared sample (which are not presented
here) give the remaining \( 6 \) optical mode temperatures. The obtained
Debye and Einstein temperatures and the belonging grouping parameters
\( \nu _{i} \) are summarized in Tab. \ref{Tabelle - Gitterfitparameter}.
The phonon energies are in good agreement with recent calculations.
\cite{Ignatov03}

\begin{figure}
{\centering \resizebox*{0.48\textwidth}{!}{\includegraphics{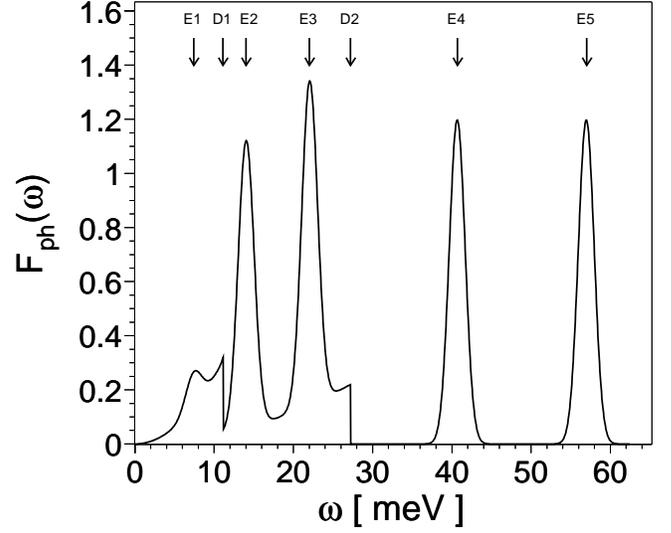}} \par}

\caption{Schematic phonon model spectrum \protect\( F_{\textrm{ph}}(\omega )
\protect \)
for \protect\( \textrm{MgC}_{1.6}\textrm{Ni}_{3}\protect \) derived
from fit parameters according to Tab. \ref{Tabelle - Gitterfitparameter}.
The peak width of the optical modes was chosen arbitrarily as \protect\( 
\sigma _{\textrm{i}}^{2}=1\protect \).\label{Fig - Phonon-Spektrum} }
\end{figure}

\begin{figure}
{\centering \resizebox*{0.4\textwidth}{!}{\includegraphics{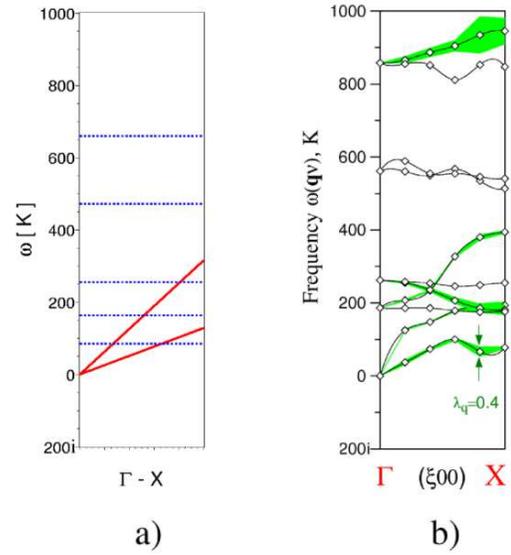}} \par}

\caption{Phonon dispersion along \protect\( \Gamma \protect \)-X direction
in the first Brillouin zone. a) Our model with acoustic phonons (black
lines) and optic phonons (dashed lines) according to Tab. 
\ref{Tabelle - Gitterfitparameter},
b) Calculations reported by \citeauthor{Ignatov03} after Ref. 
\onlinecite{Ignatov03}.\label{Fig-phonondispersion}}
\end{figure}

The corresponding phonon spectrum has the form\cite{srivastava90}
\begin{eqnarray*}
F_{\textrm{ph}}(\omega ) & = & 3\omega ^{2}\left[ \nu _{\textrm{D}1}
\frac{\theta \left( \Omega _{\textrm{D}1}-\omega \right) }
{\Omega ^{3}_{\textrm{D}1}}+\nu _{\textrm{D}2}\frac{\theta 
\left( \Omega _{\textrm{D}2}-\omega \right) }{\Omega ^{3}_{
\textrm{D}2}}\right] +\\
 &  & +\sum ^{5}_{i=1}\frac{\nu _{\textrm{E}i}}{\sqrt{2\pi 
\sigma ^{2}_{i}}}\exp \left[ -\frac{\left( \omega -\Omega _{
\textrm{E}i}\right) ^{2}}{2\sigma ^{2}_{i}}\right] ,
\end{eqnarray*}
where \( \theta (\textrm{x}) \) is the well known step-function and
\( \Omega  \) denotes the corresponding cut-off temperatures in 
\( \textrm{meV} \).The
result including higher optical modes is shown in Fig. 
\ref{Fig - Phonon-Spektrum}.
Our model parameters even reproduce the rather complex phonon dispersion
along the \( \Gamma  \)-X direction in the first Brillouin zone at
low phonon energies, as can be seen from Fig. \ref{Fig-phonondispersion},
where the used model is compared with calculations reported by 
\citeauthor{Ignatov03}\cite{Ignatov03}
Even though our model only involves constant and linear dispersion
by the Einstein and Debye model, respectively, the calculated phonon
dispersion (right panel) is well reproduced (left panel), by means
of superpositions of acoustic and optic phonon modes. The high-energy
optic phonons obtained from the model are shifted to lower energies
than predicted by the calculations. The shift is most probably caused
by anharmonic effects, which usually increase specific heat data at
higher temperatures.\cite{Ignatov03}

To investigate the electron-phonon coupling strength, the electron-p
\textcolor{black}{honon
interaction function \( \alpha ^{2}F_{\textrm{ph}}(\omega ) \) is
of interest. The coupling function \( \alpha ^{2}(\omega ) \) is
usually extracted from tunneling measurements. In the case of A15
compounds\cite{junod83} and some borocarbides,\cite{manalo01} \( 
\alpha ^{2}(\omega ) \)
is found to be of the form \( \alpha ^{2}(\omega )=\delta /\sqrt{
\omega } \),
with a scaling parameter \( \delta  \). Within this approach the
logarithmically averaged mean phonon frequency \( \omega _{
\textrm{ln}} \)
was determined from\begin{eqnarray}
\omega _{\textrm{ln}} & = & \exp \left( \frac{2}{\lambda _{
\textrm{ph}}}\int ^{\infty }_{0}\textrm{d}\omega \frac{\alpha 
^{2}(\omega )F(\omega )}{\omega }\ln \omega \right) ,\nonumber \\
\lambda _{\textrm{ph}} & = & 2\int ^{\infty }_{0}\textrm{d}
\omega \frac{\alpha ^{2}(\omega )F(\omega )}{\omega }
\label{Eq - coupling-interaction-function} 
\end{eqnarray}
as \( \omega _{\textrm{ln}}=143\textrm{ K} \). This frequency is
used in the well known McMillan formu}la (refined by \citeauthor{allen75})
\cite{allen75}\begin{equation}
\label{Eq - Allen-Dynes}
T_{\textrm{c}}\approx \frac{\omega _{\textrm{ln}}}{1.2}\exp 
\left[ -\frac{1+\lambda _{\textrm{ph}}}{\lambda _{\textrm{ph}}-\mu ^{
\star }\left( 1+0.6\lambda _{\textrm{ph}}\right) }\right] 
\end{equation}
to estimate the electron-phonon coupling co\textcolor{black}{nstant
\( \lambda _{\textrm{ph}} \). \( \mu ^{\star } \) denotes the usually
weak Coulomb pseudopotential which has been chosen as \( \mu ^{\star }=0.13 \)
in this case. With \( T_{\textrm{c}}=6.8\textrm{ K} \) the electron-phonon
coupling constant amounts \( \lambda _{\textrm{ph}}=0.84 \), suggesting
moderate coupling as proposed, for instance in Refs. \onlinecite{he01}
and \onlinecite{lin03}. However, the low value of \( \lambda _{\textrm{ph}} \)
estimated from Eq. (\ref{Eq - Allen-Dynes}) is in contradiction with
our specific heat data as already mentioned in Sec. \ref{Sec - cp results}.
In particular, \( \lambda _{\textrm{ph}}=1.45 \) was derived from
the ratio \( \gamma _{\textrm{N}}^{\star }/\gamma _{0} \) and also
the high value of the superconducting jump \( \Delta 
\textrm{c}(T_{\textrm{c}})/(\gamma _{\textrm{N}}T_{\textrm{c}})=2.09 \)
indicates strong electron-phonon coupling. Strong electron-phonon
coupling was also predicted by \citeauthor{Ignatov03}\cite{Ignatov03}
(\( \lambda _{\textrm{ph}}=1.51 \)) on the base of the calculations
mentioned above. }

\begin{figure}
{\centering \resizebox*{0.48\textwidth}{!}{\includegraphics{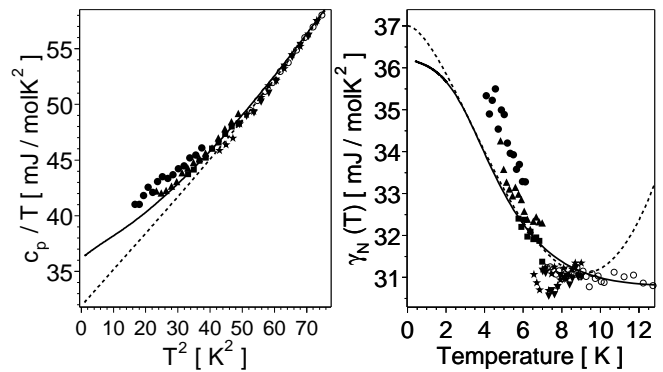}} \par}

\caption{Low temperature normal state total and electronic specific heat 
including
field measurements (\protect\( 0.5\textrm{ T}\protect \) (\textcolor{black}{
\scriptsize $\blacktriangledown$}),
\protect\( 2\textrm{ T}\protect \) (\textcolor{black}{\scriptsize \ding{72}}),
\protect\( 4\textrm{ T}\protect \) (\textcolor{black}{\tiny $\blacksquare$}),
\protect\( 6\textrm{ T}\protect \) (\textcolor{black}{\scriptsize $
\blacktriangle$})
and \protect\( 8\textrm{ T}\protect \) (\textcolor{black}{\tiny \ding{108}})).
Left panel: Specific heat data \protect\( \textrm{c}_{\textrm{p}}(T)/T
\protect \)
vs. \protect\( T^{2}\protect \). Dotted line: Extended lattice model
describing the zero field data (see Fig. \ref{Bild - spezifische 
W=E4rme H=3D0}).
Solid line: Fit of the model (including lattice and paramagnon contribution)
to the data. Right panel: electronic specific heat \protect\( 
\gamma _{\textrm{N}}(T)\protect \)
vs. \protect\( T\protect \) in the normal state. Black line: Sommerfeld
parameter \protect\( \gamma _{\textrm{N}}(T)\protect \) of the model
(see text for details), describing the observed upturn of the specific
heat at low temperatures. Dotted line: Qualitative model for spin
fluctuations according to Eq. (\ref{Eq - Paramagnon-Approximation}).
\label{Bild - spezifische W=E4rme H=3D0 inset}}
\end{figure}

\textcolor{black}{In this context a more precise analysis of the low
temperate normal state specific heat data is required. As can be seen
from the dashed line in Fig. \ref{Bild - spezifische W=E4rme H=3D0 inset},
the extended lattice model does not describe the magnetic field data.
Even larger deviations are observed if the experimental data are described
within the low temperature limit of the Debye model (see Fig. 
\ref{Bild - spezifische W=E4rme komplett}).
\citeauthor{lin03}\cite{lin03} who found a similar upturn of the
experimental data at low temperatures tried to explain this behavior
by the presence of Ni impurities. However, our x-ray analysis (see
Fig. \ref{xray Diffraktogramm}) shows no indication for Ni impurities
in our sample. Recently, \citeauthor{shan03}\cite{shan03} found that
the mentioned upturn can be easily reduced by lowering the carbon
content. They attributed the observed upturn to some kind of boson
mediated electron-electron interactions in \( 
\textrm{MgC}_{\textrm{x}}\textrm{Ni}_{3} \).
This argument motivated us to search for other possible sources to
explain the low temperature upturn of the normal state specific heat
data.}

\textcolor{black}{The easiest explanation is an additional electron-boson
interaction which may be an}

\begin{enumerate}
\item \textcolor{black}{electron-phonon interaction originating from additional
phonon-soften}ing of the lowest acoustic mode (suggested by 
\citeauthor{Ignatov03}\cite{Ignatov03}
and verified experimentally by \citeauthor{heid03}\cite{heid03})
and / or
\item electron-paramagnon interaction (see Sec. \ref{Sec - Introduction}).
\end{enumerate}
\textcolor{black}{Specific heat measurements let not clearly distinguish
between these possible origins, but since magnetization measurements
on our sample (not presented here) show increasing spin fluctuations
below \( \sim 30\textrm{ K} \) in accord with previous statements
(see Sec. \ref{Sec - Introduction}), the focus in this paper lies
on the electron-paramagnon interaction scenario. This is additionally
supported by a small magnetic field dependence of the specific heat
data, typically found in the presence of ferromagnetic spin fluctuations.}

\textcolor{black}{Within Eliashberg theory the renormalized normal
state specific heat is described by the temperature dependent thermal
mass \( \Delta m^{\star }\left( T\right) /m_{\textrm{band}} \). Its
contribution to the specific heat is given by\[
\Delta \gamma _{\textrm{sf}}(T)=\frac{\Delta m^{\star }
\left( T\right) }{m_{\textrm{band}}}\gamma _{0},\]
with\cite{rainer86}}

\textcolor{black}{\begin{eqnarray*}
\frac{\Delta m^{\star }\left( T\right) }{m_{\textrm{band}}} & = & 
\frac{6}{\pi k_{\textrm{B}}T}\int ^{\infty }_{0}\textrm{d}\omega 
\alpha ^{2}F(\omega )\left\{ -z\right. \\
 & \qquad - & \left. 2z^{2}\textrm{Im}\left[ \psi ^{\prime }(iz)\right] 
-z^{3}\textrm{Re}\left[ \psi ^{\prime \prime }(iz)\right] \right\} ,
\end{eqnarray*}
where \( \psi (iz) \) is the digamma function and \( z=\omega /
\left( 2\pi k_{\textrm{B}}T\right)  \).
The additional electron-paramagnon interaction function} is of the
form\begin{equation}
\label{Eq - el-sf interaction function}
\alpha ^{2}F_{\textrm{sf}}=a\omega \theta \left( 
\Omega _{\textrm{P}}-\omega \right) +\frac{b}{\omega ^{3}}
\theta \left( \omega -\Omega _{\textrm{P}}\right) .
\end{equation}
A corresponding fit results in a paramag\textcolor{black}{non-model
temperatu}re of \( \Omega _{\textrm{P}}\approx 2.15\textrm{ meV}
\Rightarrow \Theta _{\textrm{P}}\approx 25\textrm{ K} \)
with a thermal mass of \( \Delta m^{\star }(T=0\textrm{ K})/m\approx 0.43 \),
which is of the same order of magnitude as determined by 
\citeauthor{shan03}\cite{shan03}.
Since this low energy excitation concerns the electronic part of the
specific heat, we add it to the Sommerfeld parameter which than becomes
temperature dependent. The electronic contribution to the specific
heat increases from initially \( \gamma ^{\star }_{\textrm{N}}=31.4
\textrm{ mJ}/\textrm{molK}^{2} \)
to \( \gamma _{\textrm{N}}\left( 0\right) =36.0\textrm{ mJ}/
\textrm{molK}^{2} \).
This is understandable since the paramagnon interac\textcolor{black}{tion
dominates in the temperature range below \( 10\textrm{ K} \). The
black line in Fig. \ref{Bild - spezifische W=E4rme H=3D0 inset} shows
the good agreement of this extended model with the experimental data
in the low temperature region. The magnetic field dependence of the
paramagnons (which in addition can be temperature-dependent) is not
included in the model. In the following \( \gamma _{\textrm{N}}(T) \)
is denoted as \( \gamma _{\textrm{N}} \) for the sake of simplicity. }

\textcolor{black}{The usually applied model\begin{equation}
\label{Eq - Paramagnon-Approximation}
\gamma _{\textrm{N}}\propto \delta T^{2}\ln \left( T/T_{0}\right) ,
\end{equation}
to describe spin fluctuation behavior is shown in the right panel
of Fig. \ref{Bild - spezifische W=E4rme H=3D0 inset} for comparison. }

\begin{figure}
{\centering \textcolor{blue}{\resizebox*{0.48\textwidth}{!}{
\includegraphics{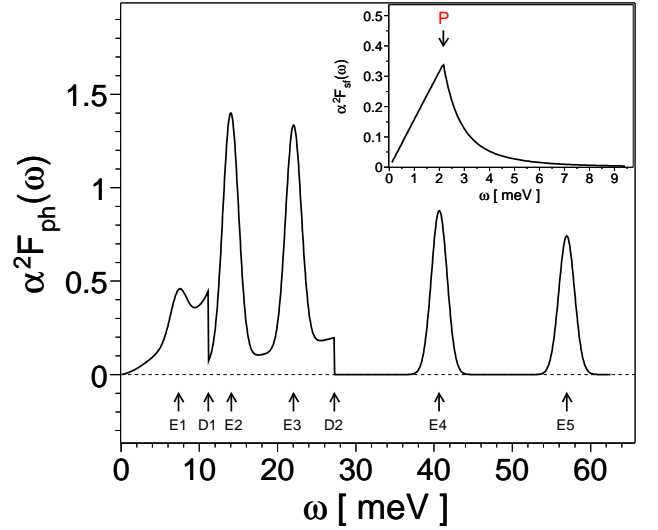}} }\par}

\caption{Electron-phonon interaction function \protect\( \alpha ^{2}F_{
\textrm{ph}}(\omega )\protect \)
for \protect\( \textrm{MgC}_{1.6}\textrm{Ni}_{3}\protect \). Phonon
energies are marked by {}``\protect\( \textrm{E}i\protect \)'',
respectively {}``\protect\( \textrm{D}i\protect \)'' (see Fig.
\ref{Fig - Phonon-Spektrum} and Tab. \ref{Tabelle - Gitterfitparameter}).
Inset: Electron-paramagnon interaction function \protect\( \alpha ^{2}F_{
\textrm{sf}}(\omega )\protect \)
according to Eq. (\ref{Eq - el-sf interaction function}), the paramagnon
energy is marked by {}``\textcolor{red}{P}''. 
\label{Fig - interaction-functions} }
\end{figure}

\textcolor{black}{At this point the question of the strength of the
coupling may be rechecked. Including the additional electron-paramagnon
interaction, the Allen-Dynes formula Eq. (\ref{Eq - Allen-Dynes})
becomes\begin{equation}
\label{Eq - Allen-Dynes-sf}
T_{\textrm{c}}\approx \frac{\omega _{\textrm{ln}}}{1.2}\exp 
\left[ -\frac{1+\lambda }{\lambda _{\textrm{ph}}-\lambda _{
\textrm{sf}}-\mu ^{\star }\left( 1+0.6\lambda _{\textrm{ph}}
\right) }\right] ,
\end{equation}
with \( \lambda =\lambda _{\textrm{ph}}+\lambda _{\textrm{sf}} \).
Using \( \omega _{\textrm{ln}}=143\textrm{ K} \), \( T_{\textrm{c}}=6.8
\textrm{ K} \),
\( \mu ^{\star }=0.13 \) and \( \lambda _{\textrm{sf}}=0.43 \),
the electron-phonon coupling constant rises to \( \lambda _{
\textrm{ph}}=1.85 \).
Using this value, the electron-phonon interaction function based on
the approach \( \alpha (\omega )=\delta /\sqrt{\omega } \) can now
be determined by scaling the factor \( \delta  \) according to 
Eq. (\ref{Eq - coupling-interaction-function}). The electron-boson interaction
functions \( \alpha ^{2}F_{\textrm{ph}}(\omega ) \) and \( \alpha ^{2}F_{
\textrm{sf}}(\omega ) \)
are shown in Fig. \ref{Fig - interaction-functions}. }

\textcolor{black}{The reliability of the model approach for the electron-phonon
coupling function \( \alpha (\omega )=\delta /\sqrt{\omega } \) can
directly be checked from the band structure, using the ratio between
the Sommerfeld parameter \( \gamma _{\textrm{N}}(0)=36.0\textrm{ mJ}/
\textrm{molK}^{2} \)
and the free electron parameter \( \gamma _{0}=\pi ^{2}k^{2}_{
\textrm{B}}N(E_{\textrm{F}})/3=11.0\textrm{ mJ}/\textrm{molK}^{2} \),
\begin{equation}
\label{Eq - Sommerfeld}
\frac{\gamma _{\textrm{N}}(0)}{\gamma _{0}}=\left( 1+\lambda _{
\textrm{ph}}+\lambda _{\textrm{sf}}\right) .
\end{equation}
With \( \lambda _{\textrm{sf}}\approx 0.43 \), the electron-phonon
coupling constant becomes \( \lambda _{\textrm{ph}}\approx 1.9 \),
showing very good agreement between both approaches.}

\textcolor{black}{In the next section, the analysis of the specific
heat in the normal state will be extended to the superconducting state.}

\subsection{Specific heat in the superconducting state
\label{Sec - sl state analysis}}

\begin{figure}
{\centering \resizebox*{0.48\textwidth}{!}{\includegraphics{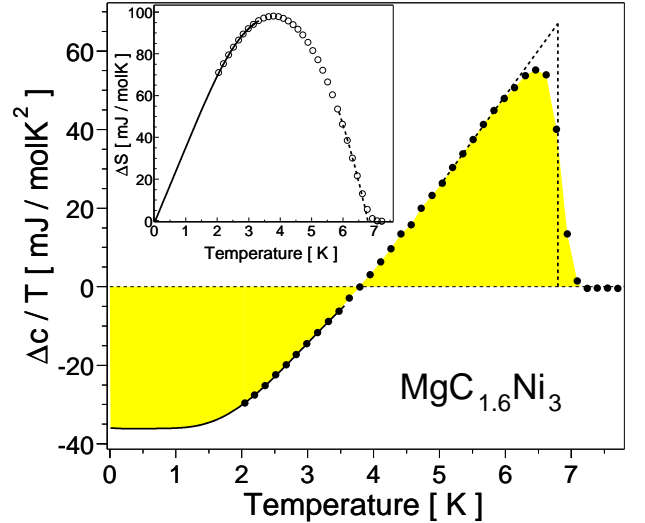}} \par}

\caption{Electronic specific heat data \protect\( \Delta \textrm{c}/T
\protect \)
vs. \protect\( T\protect \) in the superconducting state (filled
circles). The solid line in the temperature range \protect\( 0<T<3.4
\textrm{ K}\protect \)
corresponds to Eq. (\ref{Formel - BCS 1}). Dotted line: Entropy conserving
construction to get the idealized jump. Inset: Entropy conservation
for the electronic specific heat in the temperature range 
\protect\( 0<T<T_{\textrm{c}}\protect \).\label{Bild - Entropieerhaltung}}
\end{figure}
Fig. \ref{Bild - Entropieerhaltung} shows the superconducting par
\textcolor{black}{t
of the electronic specific heat \( \Delta \textrm{c}(T)=\textrm{c}_{
\textrm{p}}(T)-\textrm{c}_{\textrm{n}}(T) \),
obtained from the zero-field data. The superconducting transition
temperature \( T_{\textrm{c}}=6.8\textrm{ K} \) has been estimated
by an entropy conserving construction (dashed line in Fig. 
\ref{Bild - Entropieerhaltung}).
This value agrees well with the transition temperatures \( T_{
\textrm{c}}=6.9\textrm{ K} \)
and \( T_{\textrm{c}}=7.0\textrm{ K} \), derived from resistance
and from ac susceptibility data, respectively. The conservation of
entropy, \( \Delta S(T)=\int ^{T_{\textrm{c}}}_{0}(\Delta \textrm{c}/T)
\textrm{dT} \),
is shown in the inset of Fig. \ref{Bild - Entropieerhaltung}. It
was already mentioned, that the high value of the jump \( \Delta 
\textrm{c}(T_{\textrm{c}})/(\gamma _{\textrm{N}}T_{\textrm{c}})=2.09 \)
found for the investigated sample can be explained by strong electron-phonon
coupling. Nevertheless, we will start to analyze \( \Delta \textrm{c}(T) \)
for \( T<T_{\textrm{c}}/2 \) within the BCS theory, since the deviation
from the weak-coupling temperature dependence of the gap is mainly
restricted to the vicinity of the jump. The temperature dependence
of \( \Delta \textrm{c}(T)=\textrm{c}_{\textrm{p}}(T)-\textrm{c}_{
\textrm{n}}(T) \)
in the weak-coupling BCS theory (\( T_{\textrm{c}}\ll \omega _{\textrm{ln}} \))
is given by an approximative formula\begin{equation}
\label{Formel - BCS 1}
\Delta \textrm{c}(T)=8.5\gamma _{\textrm{N}}T_{\textrm{c}}\exp 
\left( -0.82\frac{\Delta _{\textrm{BCS}}(0)}{k_{\textrm{B}}T}\right) -
\gamma _{\textrm{N}}T,
\end{equation}
valid in the temperature range of \( 2<T_{\textrm{c}}/T<6 \) corresponding
in this case to \( 1\textrm{ K}<T<3.4\textrm{ K} \). Eq. (\ref{Formel - BCS 1})
can be fitted to the data by using the phenomenological gap \( 2\Delta _{
\textrm{exp}}/k_{\textrm{B}}T_{\textrm{c}}=3.75 \),
slightly exceeding the BCS weak coupling prediction \( 2\Delta _{
\textrm{BCS}}(0)/k_{\textrm{B}}T_{\textrm{c}}=3.52 \).
The fit, which is shown as black line in Fig. \ref{Bild - Entropieerhaltung},
describes the experimental data in the range of \( 2\textrm{ K}<T<3.5
\textrm{ K} \)
quite well.}
\begin{figure}
{\centering \resizebox*{0.48\textwidth}{!}{\includegraphics{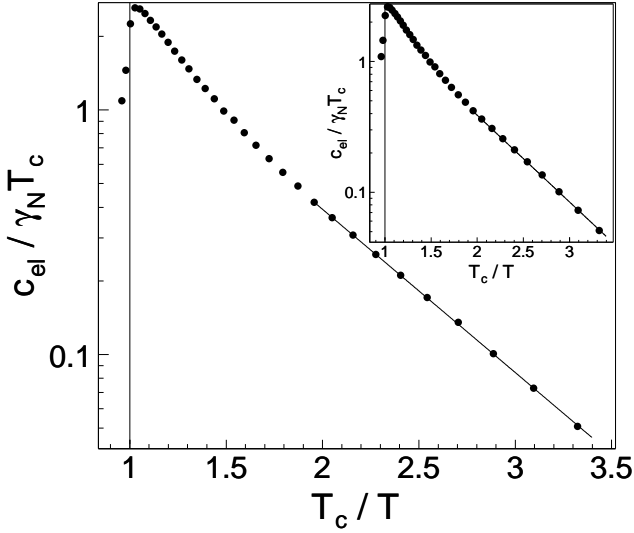}} \par}

\caption{\textcolor{black}{Normalized electronic specific heat contribution
vs. \protect\( T_{\textrm{c}}/T\protect \). The black line is a fit
of Eq. (\ref{Formel - BCS Theory}) to the experimental data. The
inset shows a fit of the two-band approximation given by 
Eq. (\ref{Formel - BCS twoband}).\label{Bild BCS - Thermodynamik}}}
\end{figure}

To examine the temperature dependence of the electronic specific he
\textcolor{black}{at\begin{eqnarray*}
\textrm{c}_{\textrm{el}} & = & \textrm{c}_{\textrm{p}}(T)-\textrm{c}_{
\textrm{lattice}}(T)\label{Formel - c_el} \\
 & = & \Delta \textrm{c}(T)+\gamma _{\textrm{N}}T_{\textrm{c}}\nonumber 
\end{eqnarray*}
at \( H=0 \) in detail, \( \textrm{c}_{\textrm{el}}(T)/\gamma _{
\textrm{N}}T_{\textrm{c}} \)
is plotted logarithmically vs. \( T_{\textrm{c}}/T \) (Fig. 
\ref{Bild BCS - Thermodynamik}).
The corresponding formula to Eq. (\ref{Formel - BCS 1}) reads
\begin{equation}
\label{Formel - BCS Theory}
\frac{\textrm{c}_{\textrm{el}}(T)}{\gamma _{\textrm{N}}T_{
\textrm{c}}}=8.5\exp \left( -0.82\frac{\Delta _{\textrm{exp}}}{k_{
\textrm{B}}T}\right) ,
\end{equation}
if \( \Delta _{\textrm{BCS}}(0) \) is replaced by \( \Delta _{
\textrm{exp}} \).
The black line in Fig. \ref{Bild BCS - Thermodynamik} is a fit of
Eq. (\ref{Formel - BCS Theory}) to the experimental data, which show
an exponential temperature dependence at low temperatures \( (T_{
\textrm{c}}/T\geq 2) \).
This is a strong indication for} \textcolor{black}{\emph{s}}
\textcolor{black}{-wave
superconductivity in \( \textrm{MgC}_{1.6}\textrm{Ni}_{3} \). A natural
explanation for the discrepancy from the expected BCS gap value of
\( 2\Delta _{\textrm{BCS}}(0)/k_{\textrm{B}}T_{\textrm{c}}=3.52 \)
to the experimentally found \( 2\Delta _{\textrm{exp}}/k_{\textrm{B}}T_{
\textrm{c}}=3.75 \)
emerges from}

\begin{enumerate}
\item \textcolor{black}{the two-band character of \( \textrm{MgC}_{1.6}
\textrm{Ni}_{3} \)
and}
\item \textcolor{black}{the enhanced electron-phonon coupling.}
\end{enumerate}
\textcolor{black}{The electronic specific heat can be analyzed within
a two-gap model, simply by extending Eq. (\ref{Formel - BCS Theory})
with a second gap\begin{eqnarray}
\frac{\textrm{c}_{\textrm{el}}(T)}{\gamma _{\textrm{N}}T_{
\textrm{c}}} & = & 8.5\left[ 0.85\exp \left( -0.82\frac{\Delta _{1}}{k_{
\textrm{B}}T}\right) \right. \nonumber \\
 & \qquad + & \left. 0.15\exp \left( -0.82\frac{\Delta _{2}}{k_{
\textrm{B}}T}\right) \right] ,\label{Formel - BCS twoband} 
\end{eqnarray}
using \( 85\textrm{ }\% \) contribution for the hole band and \( 15
\textrm{ }\% \)
contribution for the electron band. The dotted line in Fig. 
\ref{Bild BCS - Thermodynamik}
shows this fit for two gaps with \( 2\Delta _{1}/k_{\textrm{B}}T_{
\textrm{c}}=3.67 \)
(fitted parameter) and \( 2\Delta _{2}/k_{\textrm{B}}T_{\textrm{c}}=4.50 \)
(fixed parameter). This situation is nearly indistinguishable from
the single band model (see inset of Fig. \ref{Bild BCS - Thermodynamik}).
The two-gap model is of considerable interest, since in this case
the large gap found in recent tunneling measurements of Ref. \onlinecite{mao03}
(\( 2\Delta /k_{\textrm{B}}T_{\textrm{c}}=4.6 \)) and 
Ref. \onlinecite{shan03pc}
(\( 2\Delta /k_{\textrm{B}}T_{\textrm{c}}=4.3 \)) is reproduced. }

\textcolor{black}{Even the lower gap of the two-gap model exceeds
the BCS weak-coupling limit. This and the strongly enhanced specific
heat jump \( \Delta \textrm{c}(T_{\textrm{c}}) \) are clear indications
of strong electron-phonon coupling in accordance with our normal state
specific heat analysis. Thus it is now straightforward to investigate
the electron-phonon coupling strength and thus, the characteristic
phonon frequency \( \omega _{\textrm{ln}} \), introduced in Sec.
\ref{Sec - normal state cp analysis}, from the superconducting state
characteristics. The Eliashberg theory provides the following approximate
formulas, which includes strong coupling corrections within an isotropic
single band model and links \( x=\omega _{\textrm{ln}}/T_{\textrm{c}} \)
to experimental thermodynamic quantities:\cite{carbotte90}
\begin{subequations}\label{carbotte:alle}\begin{eqnarray}
\frac{2\Delta (0)}{k_{\textrm{B}}T_{\textrm{c}}} & = & 3.53B_{0}
\left( x\right) ,\label{carbotte0} \\
\frac{\Delta \textrm{c}(T_{\textrm{c}})}{\gamma _{\textrm{N}}T_{
\textrm{c}}} & = & 1.43B_{1}\left( x\right) ,\label{carbotte1} \\
\frac{\Delta \textrm{c}(T)-\Delta \textrm{c}(T_{\textrm{c}})}{
\gamma _{\textrm{N}}T_{\textrm{c}}-\gamma _{\textrm{N}}T} & = & -3.77B_{2}
\left( x\right) ,\label{carbotte2} \\
\frac{\gamma _{\textrm{N}}T^{2}_{\textrm{c}}}{H^{2}_{
\textrm{c}}(0)} & = & 0.168B_{3}\left( x\right) ,\label{carbotte3} \\
\frac{H_{\textrm{c}}(0)}{\left. \frac{\textrm{d}H_{\textrm{c}}}{
\textrm{d}T}\right| _{T_{\textrm{c}}}T_{\textrm{c}}} & = & 0.576B_{4}
\left( x\right) .\label{carbotte4} 
\end{eqnarray}
\end{subequations}The corresponding logarithmic correction terms are
given by\begin{subequations}\label{carbotte:alleb}\begin{eqnarray}
B_{0}(x) & = & 1+12.5x^{-2}\ln \frac{x}{2},\label{carbotte0b} \\
B_{1}(x) & = & 1+53x^{-2}\ln \frac{x}{3},\label{carbotte1b} \\
B_{2}(x) & = & 1+117x^{-2}\ln \frac{x}{2.9},\label{carbotte2b} \\
B_{3}(x) & = & 1-12.2x^{-2}\ln \frac{x}{3},\label{carbotte3b} \\
B_{4}(x) & = & 1-13.4x^{-2}\ln \frac{x}{3.5}.\label{carbotte4b} 
\end{eqnarray}
\end{subequations}}

\textcolor{black}{Now, using Eq. (\ref{carbotte0}), \( T_{
\textrm{c}}=6.8\textrm{ K} \)
and the gap value \( \Delta _{\textrm{exp}}(2\textrm{ K})=1.10\textrm{ meV} \)
, one arrives at \( \omega _{\textrm{ln}}=149\textrm{ K} \).}

\textcolor{black}{Using the value of the idealized jump of the specific
heat, \( \Delta \textrm{c}(T_{\textrm{c}})/\left( \gamma _{
\textrm{N}}T_{\textrm{c}}\right) =2.09 \)
in Eq. (\ref{carbotte1}) with \( T_{\textrm{c}}=6.8\textrm{ K} \),
\( \omega _{\textrm{ln}}=88\textrm{ K} \) is derived.}

\textcolor{black}{Comparing the linear slope of the idealized specific
heat in the superconducting state of \( -6.7 \), obtained from Fig.
\ref{Bild - Entropieerhaltung} with Eq. (\ref{carbotte2}), one gets
\( \omega _{\textrm{ln}}=109\textrm{ K} \). }
\begin{figure}
{\centering \resizebox*{0.48\textwidth}{!}{\includegraphics{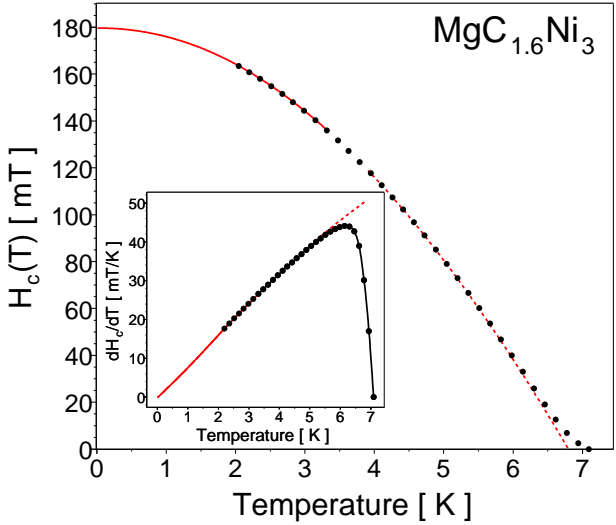}} \par}

\caption{Temperature dependence of the thermodynamic critical field 
\protect\( H_{\textrm{c}}(T)\protect \)
(filled circles) derived from the electronic specific heat in the
superconducting state using Eq. (\ref{Formel - Hcthermodynamic}).
Solid line (\protect\( 0<T<3.4\textrm{ K}\protect \)): Single band
model according to Eq. (\ref{Formel - BCS Theory}). Dotted line:
Idealized jump construction (see Fig. \ref{Bild BCS - Thermodynamik}).
Inset: Derivative \protect\( \textrm{d}H_{\textrm{c}}/\textrm{d}T\protect \)
vs. \protect\( T\protect \) (filled circles) and idealized jump (dotted
line).\textcolor{red}{\label{Bild - Hcthermodynamic}}}
\end{figure}

\textcolor{black}{In view of strong-coupling effects the ratio 
\( \gamma _{\textrm{N}}T^{2}_{\textrm{c}}/H^{2}_{\textrm{c}}(0) \),
implying again only thermodynamic quantities, is of interest. Known
superconductors show values between \( 0.17\ldots 0.12 \) ranging
from weak to strong coupling, respectively (see for example page 1086
of Ref. \onlinecite{carbotte90}). The thermodynamic critical field
\( H_{\textrm{c}}(T) \) can be determined with the help of the Gibbs
free energy \( \textrm{d}F=-S\textrm{d}T-M\textrm{d}B \) as\begin{equation}
\label{Formel - Hcthermodynamic}
H_{\textrm{c}}(T)=\sqrt{-8\pi \Delta F}.
\end{equation}
\( \Delta F \) is to be extracted from the specific heat in the superconducting
state, \( \Delta \textrm{c}(T)=-T\textrm{d}^{2}(\Delta F)/\textrm{d}T^{2} \)}
\textcolor{blue}{.}
\textcolor{black}{The temperature dependence of \( H_{\textrm{c}}(T) \)
is shown in Fig. \ref{Bild - Hcthermodynamic}. With \( H_{\textrm{c}}(0)=179.6
\textrm{ mT} \)
we found \( \gamma _{\textrm{N}}T^{2}_{\textrm{c}}/H^{2}_{
\textrm{c}}(0)=0.155 \).
From Eq. (\ref{carbotte3}) we get \( \omega _{\textrm{ln}}=110\textrm{ K} \)
(with \( T_{\textrm{c}}=6.8\textrm{ K} \)).}

\textcolor{black}{Next, from the derivative of the thermodynamic critical
field at zero temperature, \( \textrm{d}H_{\textrm{c}}/\textrm{d}T \)
the ratio \( H_{\textrm{c}}(0)/\left( \left. \textrm{d}H_{\textrm{c}}(T)/
\textrm{d}T\right| _{T_{\textrm{c}}}T_{\textrm{c}}\right)  \)
can be estimated. The value at \( T=T_{\textrm{c}} \) (of the idealized
jump construction), amounts \( \left. \textrm{d}H_{\textrm{c}}(T)/
\textrm{d}T\right| _{T_{\textrm{c}}}=50.236 \)
(see dashed line in the inset of Fig. \ref{Bild - Hcthermodynamic}).
Using the experimental value of \( H_{\textrm{c}}(0)/\left( \left. 
\textrm{d}H_{\textrm{c}}(T)/\textrm{d}T\right| _{T_{\textrm{c}}}T_{
\textrm{c}}\right) =0.525 \)
in Eq. (\ref{carbotte4}), a value of \( \omega _{\textrm{ln}}=102\textrm{ K} \)
is extracted.}

\textcolor{black}{It should be noted that Eqs. (\ref{carbotte0})
and (\ref{carbotte4}) can be used to estimate the value of the gap
\( \Delta (0) \) from the thermodynamic critical field \( H_{\textrm{c}}(0) \),
due to similar dependences on strong coupling corrections:\cite{gladstone}\[
\left. \left( \frac{T}{H_{\textrm{c}}(0)}\frac{\textrm{d}H_{\textrm{c}}(T)}{
\textrm{d}T}\right) \right| _{T=T_{\textrm{c}}}\approx \frac{\Delta (0)}{k_{
\textrm{B}}T_{\textrm{c}}}.\]
Using \( \left. \textrm{d}H_{\textrm{c}}(T)/\textrm{d}T\right| _{T_{
\textrm{c}}}=50.236 \),
we get \( 2\Delta (0)/k_{\textrm{B}}T_{\textrm{c}}\approx 3.80 \),
agreeing well with the single band result \( 2\Delta _{\textrm{exp}}/k_{
\textrm{B}}T_{\textrm{c}}=3.75 \)
of Eq. (\ref{Formel - BCS Theory}).}

\begin{figure}
{\centering \resizebox*{0.48\textwidth}{!}{\includegraphics{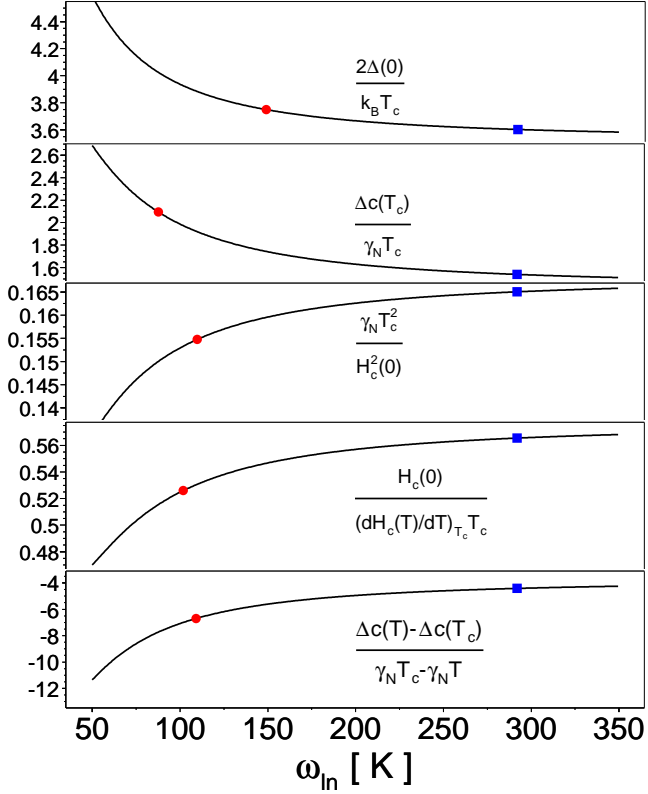}} \par}

\caption{Several thermodynamic quantities in depen\textcolor{black}{dence
on the charac}teristic phonon frequency \protect\( \omega _{\textrm{ln}}
\protect \)
according to Eqs. (\ref{carbotte:alle}) and (\ref{carbotte:alleb}).
Filled circles: thermodynamic quantities estimated for \protect\( 
\textrm{MgC}_{1.6}\textrm{Ni}_{3}\protect \)
from experimental data. Strong discrepancies are found within the
low temperature Debye limit (filled squares). Note that the weak-coupling
limit is reached in the asymptotic extrapolation \textcolor{black}{
\protect\( \omega _{\textrm{ln}}\rightarrow \infty \protect \).}
\label{Bild - Results}}
\end{figure}

\textcolor{black}{In summary, \( \omega _{\textrm{ln}} \) was estimated
from five different thermodynamic relations, only involving experimental
results. The mean value \( \overline{\omega _{\textrm{ln}}}=(111\pm 23)
\textrm{ K} \)
is in good agreement with calculations of \citeauthor{Ignatov03}
\cite{Ignatov03}
An overview of the results is given in Fig. \ref{Bild - Results}.
Note, that a similar analysis was already successfully used to describe
some borocarbide superconductors.\cite{michor96,manalo01}}

\textcolor{black}{The mean value \( \overline{\omega _{\textrm{ln}}} \),
derived from the superconducting state is somewhat smaller than the
normal state result, \( \omega _{\textrm{ln}}=143\textrm{ K} \).
This may be attributed to an additional phonon softening contribution
or the approximative approach of the electron-phonon coupling function
\( \alpha ^{2}(\omega ) \) (see Sec. \ref{Sec - normal state cp analysis}).
Nevertheless by checking Eq. (\ref{Eq - Allen-Dynes-sf}) with \( 
\overline{\omega _{\textrm{ln}}}=(111\pm 23)\textrm{ K} \),
\( T_{\textrm{c}}=6.8\textrm{ K} \), \( \mu ^{\star }=0.13 \) and
\( \lambda _{\textrm{sf}}=0.43 \) the electron-phonon coupling constant
becomes \( \lambda _{\textrm{ph}}=1.9\ldots 2.3 \), whereas \( \lambda _{
\textrm{ph}}=1.85 \)
was derived from \( \omega _{\textrm{ln}}=143\textrm{ K} \) for the
same parameters. It should be noted here, that Eqs. (\ref{carbotte:alle})
were derived assuming a small value for the Coulomb pseudopotential
\( \mu ^{\star } \), which is oversimplified considering enhanced
electron-paramagnon coupling found in this analysis. A rough correction
would shift the characteristic phonon frequency \( \omega _{\textrm{ln}} \)
to slightly higher values and a coupling constant of \( \lambda _{
\textrm{ph}}\approx 1.9 \)
seems to be most likely. Note that a similar analysis was already
successfully used to describe some borocarbide superconductors.
\cite{michor96,manalo01}}

\begin{figure}
{\centering \textcolor{blue}{\resizebox*{0.48\textwidth}{!}{
\includegraphics{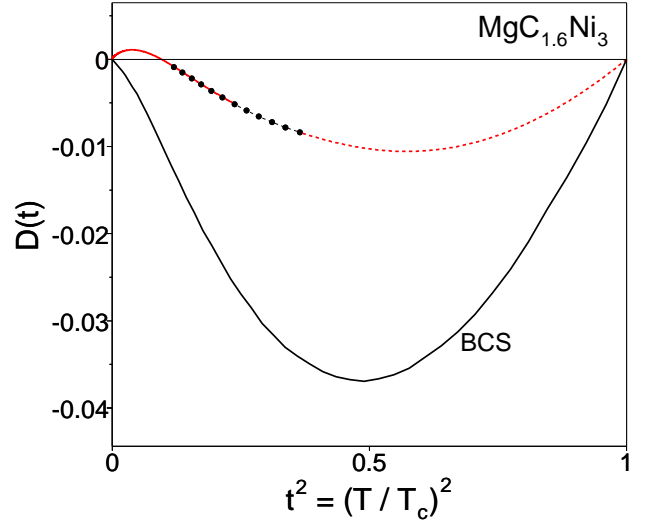}} }\par}

\caption{\textcolor{black}{Deviation function of the thermodynamic critical
field of \protect\( \textrm{MgC}_{1.6}\textrm{Ni}_{3}\protect \)
(filled circles) as function of \protect\( \left( T/T_{\textrm{c}}
\right) ^{2}\protect \).
The solid line for \protect\( 0<T<0.34\textrm{ K}\protect \) corresponds
to Eq. (\ref{Formel - BCS Theory}), the dotted line corresponds to
the idealized jump construction (see Fig. \ref{Bild - Entropieerhaltung}).
For comparison, the weak coupling BCS result\cite{muehl59} is shown.}
\label{Fig - deviation function}}
\end{figure}
The analysis of the thermodynamic properties of \( \textrm{MgCNi}_{3} \)
presented so far clearly points to strong electron-phonon coupling.
However, the temperature dependence of the thermodynamic critical
field \( H_{\textrm{c}}(T) \) shown in Fig. \ref{Bild - Hcthermodynamic}
strongly deviates from analogous data for well-known strong coupling
superconductors such as \( \textrm{Hg} \) or \( \textrm{Pb} \).
\( H_{\textrm{c}}(T) \) is usually analyzed in terms of the deviation
function \textcolor{black}{\( D(t)=H_{\textrm{c}}(T)=H_{\textrm{c}}(0)
\left( 1-t^{2}\right)  \)}
wi\textcolor{black}{th \( t=T/T_{\textrm{c}} \). T}he deviation function
of the above mentioned strong coupling superconductors is positive
and goes through a maximum at \( t^{2}\approx 0.5 \). The deviation
function of \( \textrm{MgCNi}_{3} \) is shown in Fig. 
\ref{Fig - deviation function}.
Instead of the expected positive sign\textcolor{black}{, \( D(t^{2}) \)
of \( \textrm{MgCNi}_{3} \) becomes negative already above about
\( 0.3T_{\textrm{c}} \). The shape of the deviation function of \( 
\textrm{MgCNi}_{3} \)
closely resembles that one of \( \textrm{Nb} \) having an electron-phonon
coupling strength of \( \lambda _{\textrm{ph}}\approx 1.0 \). We
remind the reader that the weak coupling BCS model yields a negative
maximum deviation of \( \approx 3.8\textrm{ }\% \) (see Fig. 
\ref{Fig - deviation function}).
Thus, at first glance, our result seems to be in contradiction with
the strong electron-phonon coupling suggested above. It turns out
that this contradiction can be resolved, taking into account a splitting
of the electron-phonon interaction function in a high and a low (soft)
energy part. This is illustrated in Fig. \ref{Fig - 
calculated-deviation-function},
where a two phonon peak spectrum with equal coupling strengths of
both peaks located at \( \omega _{1} \) and \( \omega _{2} \) has
been analyzed in the strong coupling case of \( \lambda _{\textrm{ph}}
\approx 2 \)
under consideration. The theoretical curves calculated within the
ISB are shown for different frequency ratios \( \omega _{1}/\omega _{2} \).
For \( \omega _{1}/\omega _{2}\approx 8 \), the {}``standard''
strong coupling behavior, namely a positive deviation function, is
co}mpletely removed and the deviation function becomes negative. Considering
the low energy modes \( \textrm{E}1 \) and \( \textrm{D}1 \), found
in the analysis of the specific heat in the normal state (see Figs.
\ref{Fig - Phonon-Spektrum} and \ref{Fig - interaction-functions}),
this situation is easily imaginable to be valid in the case of \( 
\textrm{MgCNi}_{3} \). 
\begin{figure}
{\centering \resizebox*{0.48\textwidth}{!}{\includegraphics{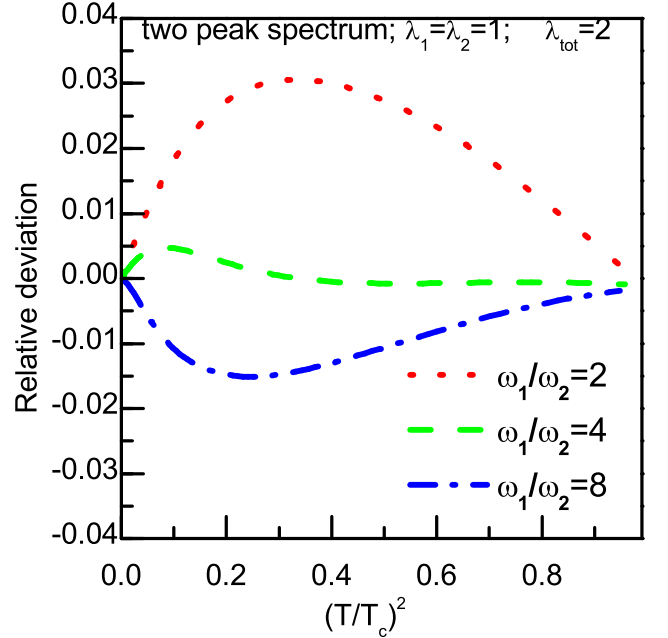}} \par}

\caption{Normalized deviation function calculated within the Eliashberg theory
for an idealized two-peak phonon spectrum located at 
\protect\( \omega _{1}\protect \)
and \protect\( \omega _{2}\protect \) with equal electron-phonon
coupling parameters \protect\( \lambda _{1}=\lambda _{2}=1\protect \)
and strong total coupling parameter of \protect\( \lambda _{
\textrm{ph},\textrm{tot}}=2\protect \).
Shown are results for \protect\( \omega _{1}/\omega _{2}=2,4
\textrm{ and }8\protect \).\label{Fig - calculated-deviation-function} }
\end{figure}

In the superconducting state a linear-in-\( T \) electronic specific
heat contribution \( \gamma (H)T \) arises from the normal conducting
cores of the flux lines for applied magnetic fields \( H>H_{\textrm{c}1} \).

\begin{figure}
{\centering \resizebox*{0.48\textwidth}{!}{\includegraphics{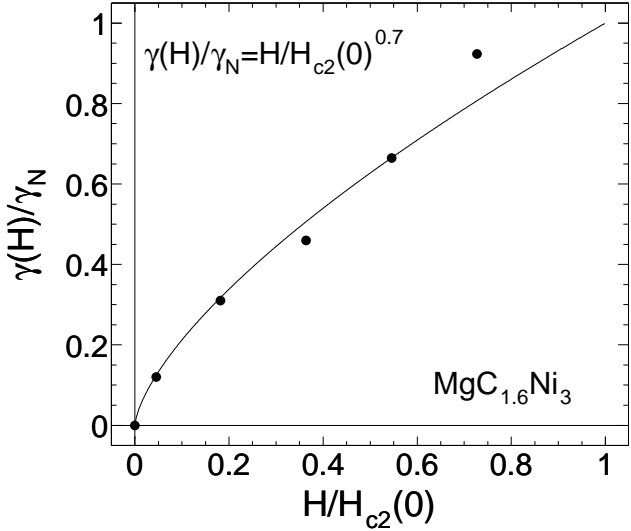}} \par}

\caption{Normalized field-dependent Sommerfeld parameter \protect\( 
\gamma (H)/\gamma _{\textrm{N}}\protect \)
plotted aga\textcolor{black}{inst \protect\( H/H_{\textrm{c}2}(0)\protect \).
Filled circles: \protect\( \gamma (H)/\gamma _{\textrm{N}}=\left[ 
\textrm{c}_{\textrm{p}}(T,H)-\textrm{c}_{\textrm{p}}(T,0)\right] /\gamma _{
\textrm{N}}\protect \)
at \protect\( T=2\textrm{ K}\protect \) for different applied magnetic
fields. The black line is a} fit of \protect\( \gamma (H)/\gamma _{
\textrm{N}}=\left( H/H_{\textrm{c}2}(0)\right) ^{0.7}\protect \)
using \protect\( H_{\textrm{c}2}(0)=11\textrm{ T}\protect \) and
\protect\( \gamma _{\textrm{N}}\protect \) at \protect\( T=2\textrm{ K}
\protect \).\label{Bild - gamma von H}}
\end{figure}

This contribution can be expressed as \( \gamma (H)T=\textrm{c}_{
\textrm{p}}(T,H)-\textrm{c}_{\textrm{p}}(T,0) \),\cite{sonier98}
where \( \textrm{c}_{\textrm{p}}(T,0) \) is the specific heat in
the Meissner state. Specific heat data for \( \textrm{MgC}_{1.6}
\textrm{Ni}_{3} \)
at \( T=2\textrm{ K} \) were analyzed in order to derive the field
dependence of \( \gamma (H) \). In Fig. \ref{Bild - gamma von H},
the obtained \( \gamma (H)/\gamma _{\textrm{N}} \) is plotted against
\( H/H_{\textrm{c}2}(0) \) using \( H_{\textrm{c}2}(0)=11.0\textrm{ T} \).

The field data of \( \textrm{c}_{\textrm{p}}/T \) shown in Fig. 
\ref{Bild - gamma von H}
can be descri\textcolor{black}{bed in accord with results from Ref.
\onlinecite{lin03} by th}e expression \( \gamma /\gamma _{\textrm{N}}=
\left( H/H_{\textrm{c}2}(0)\right) ^{0.7} \)
which differs from the linear \( \gamma (H) \) law expected for isotropic
\emph{s}-wave superconductors in the dirty limit.

A non-linear field dependence close to \( \gamma (H)\propto H^{0.5} \)
has been reported for some unconventional superconductors with gap
nodes in the quasiparticle spectrum of the vortex state as \( 
\textrm{YBa}_{2}\textrm{Cu}_{3}\textrm{O}_{7} \),\cite{wright99}
and in the heavy fermion superconductor \( \textrm{UPt}_{3} \),
\cite{ramirez95}
but also in some clean \emph{s}-wave superconductors as \( 
\textrm{CeRu}_{2} \),\cite{hedo98}
\( \textrm{NbSe}_{2} \)\cite{sonier98,nohara99} and the borocarbides
\( \textrm{RNi}_{2}\textrm{B}_{2}\textrm{C } \) \( (\textrm{R}=
\textrm{Y},\textrm{ Lu}) \).\cite{nohara97,lipp02}
Delocalized quasiparticle states around the vortex cores, similar
as in \emph{d}-wave superconductors, seem to be responsible for the
non-linear \( \gamma (H) \) dependence in the boroc\textcolor{black}
{arbides.\cite{izawa01,boaknin01}}

\subsection{\textcolor{black}{The main superconducting and 
thermodynamic parameters}}

\textcolor{black}{In this subsection we collect the values of the
main physical parameters we have found experimentally and compare
them with available data in the literature. In order to make this
comparison as complete as possible we estimate (calculate), from our
data and from those of Ref. \onlinecite{mao03}, the lower critical
field \( \textrm{H}_{\textrm{c}1}(0) \) and the penetration depth
\( \lambda _{\textrm{L}}(0) \) at zero temperature adopting the 
applicability
of the standard Ginzburg-Landau (GL) theory. Within this theory the
penetration depth is given by the relation\begin{equation}
\label{Eq - penetration depth}
\lambda _{\textrm{L}}(0)=\kappa (0)\xi _{\textrm{GL}}(0),
\end{equation}
where the Ginzburg-Landau coherence length \( \xi _{\textrm{GL}}(0) \)
and the Ginzburg-Landau parameter \( \kappa  \) are related to the
upper and the thermodynamic critical fields as:\begin{eqnarray*}
\xi _{\textrm{GL}}(0) & = & \sqrt{\Phi _{0}/2\pi H_{\textrm{c}2}(0)},\\
\kappa (0) & = & \frac{H_{\textrm{c}2}(0)}{\sqrt{2}H_{\textrm{c}}(0)},
\end{eqnarray*}
with the flux quantum \( \Phi _{0} \). With \( H_{
\textrm{c}2}(0)=11\textrm{ T} \)
and \( H_{\textrm{c}}(0)=180\textrm{ mT} \) (see Sec. 
\ref{Sec - sl state analysis}),
\( \xi _{\textrm{GL}}(0)=5.47\textrm{ nm} \) and \( 
\kappa (0)=43.3 \)
are obtained. Using theses values in Eq. (\ref{Eq - penetration depth}),
the penetration depth is estimated to be \( \lambda _{
\textrm{L}}(0)=237\textrm{ nm} \).
Our calculated value agrees well with measurements performed by 
\citeauthor{prozorov03}\cite{prozorov03}
resulting \( \lambda _{\textrm{L}}(0)=(250\pm 20)\textrm{ nm} \).
It should be noted that \citeauthor{lin03}\cite{lin03} measured a
penetration depth of \( \lambda _{\textrm{L}}(0)=(128\ldots 180)\textrm{ nm} \)
for their sample (see also Ref. \onlinecite{lin}), possible consequences
will be discussed in Sec. \ref{Sec - Hc2 analysis}. To complete the
critical field analysis, the lower critical field \( H_{\textrm{c}1}(0) \)
can be estimated using\[
H_{\textrm{c}1}(0)H_{\textrm{c}2}(0)=H^{2}_{\textrm{c}}(0)\left( \ln 
\kappa (0)+0.08\right) .\]
With \( H_{\textrm{c}2}(0)=11\textrm{ T} \) and \( \kappa (0)=43.3 \)
we get \( H_{\textrm{c}1}(0)=11.3\textrm{ mT} \), agreeing well with
\( H_{\textrm{c}1}(0)=12.6\textrm{ mT} \), measured by 
\citeauthor{jin03}\cite{jin03}
The results are shown in Tab. \ref{Tab - main parameters}, where
for comparison result of Refs. \onlinecite{mao03}, \onlinecite{lin} and
\onlinecite{jin03} have been inclu}ded. Comparing these sets one finds
a general qualitative accord.

\begin{table}

\caption{Main superconducting and thermodynamic electronic parameters for
\protect\( \textrm{MgCNi}_{3}\protect \).\label{Tab - main parameters}}

\begin{ruledtabular}

\begin{tabular}{lldddd}
&
&
\multicolumn{1}{r}{\textrm{Present work}\footnotemark[1]}&
\multicolumn{1}{r}{\textrm{Ref. \onlinecite{lin}}\footnotemark[1]}&
\multicolumn{1}{c}{\textrm{Ref. \onlinecite{mao03}}\footnotemark[2]}&
\multicolumn{1}{c}{\textrm{Ref. \onlinecite{jin03}}\footnotemark[2]}\\
\hline
\( T_{\textrm{c}} \)&
\( [\textrm{K}] \)&
6.8&
6.4&
7.63&
7.3\\
\( H_{\textrm{c}2}(0) \)&
\( [\textrm{T}] \)&
11&
11.5&
14.4&
16\\
\( H_{\textrm{c}}(0) \)&
\( [\textrm{T}] \)&
0.18&
\multicolumn{1}{c}{\textrm{0.29$\pm$0.04}}&
0.19&
0.6\\
\( H_{\textrm{c}1}(0) \)&
\( [\textrm{mT}] \)&
11.3\footnotemark[3]&
\multicolumn{1}{c}{\textrm{23$\pm$7}\footnotemark[3]}&
10.0\footnotemark[3]&
12.6\footnotemark[4]\\
\( \xi _{\textrm{GL}}(0) \)&
\( [\textrm{nm}] \)&
5.47&
5.4&
4.6&
4.5\\
\( \kappa (0) \)&
&
43.3&
\multicolumn{1}{c}{\textrm{29.0$\pm$5.0}}&
54.0&
51\\
\( \lambda _{\textrm{L}}(0) \)&
\( [\textrm{nm}] \)&
237.\footnotemark[3]&
\multicolumn{1}{c}{\textrm{154$\pm$26}\footnotemark[4]}&
248.\footnotemark[3]&
230.0\footnotemark[3]\\
\multicolumn{1}{l}{\( \gamma _{\textrm{N}} \)}&
\multicolumn{1}{l}{\( [\frac{\textrm{mJ}}{\textrm{molK}^{2}}] \)}&
31.4&
33.6&
30.1&
\\
\( \frac{\Delta \textrm{c}}{\gamma _{\textrm{N}}T_{\textrm{c}}} \)&
&
2.09&
1.97&
2.1&
\\
\( \Theta ^{\star }_{\textrm{D}} \)&
\( [\textrm{K}] \)&
292&
287&
284&
\\
\( \omega _{\textrm{ln}} \)&
\( [\textrm{K}] \)&
143&
135.\footnotemark[5]&
161.\footnotemark[5]&
\\
\end{tabular}

\end{ruledtabular}

\footnotetext[1]{Using a parabolically extrapolated \( H_{
\textrm{c}2}(0) \) value.} 

\footnotetext[2]{Using the artificial WHH estimate for \( H_{
\textrm{c}2}(0) \).}

\footnotetext[3]{Calculated.}

\footnotetext[4]{Measured.}

\footnotetext[5]{Calculated (Eq. (\ref{Eq - Allen-Dynes-sf}) using \( 
\lambda _{\textrm{ph}}=1.85 \), \( \lambda _{\textrm{sf}}=0.43 \),
\( \mu ^{\star }=0.13 \)).}
\end{table}

\section{Theoretical analysis and Discussion}

Naturally, the obtained parameter set is model dependent. In this
context even the case of relatively simple Fermi surfaces provides
a difficult task to solve the full three(four)-dimensional Eliashberg
problem with given \( v_{\textrm{F}}(\overrightarrow{k}) \) and \( 
\alpha ^{2}F(\overrightarrow{k},\overrightarrow{k^{\prime }},\omega ) \)
for all physical quantities of interest. However, the solution of
this problem can be sufficiently simplified for three practically
important cases:

\begin{enumerate}
\item the relatively simple standard isotropic single band model (ISB),
where \( v_{\textrm{F}}(\overrightarrow{k}) \) is constant and the
spectral function \( \alpha ^{2}F \) depends only on the boson (phonon)
frequency,
\item a separable anisotropic single band model which exploits the so called
first order Fermi surface harmonic approximation and 
\item \textcolor{black}{the isotropic two-band model (ITB). The latter is
a straightforward generalization of the ISB with respect to two order
parameters.}
\end{enumerate}
\textcolor{black}{Due to the present lack of single crystal samples
we will ignore the second issue. In addition, the cubic structure
of \( \textrm{MgCNi}_{3} \) suggests only weak anisotropy effects. }

\subsection{\textcolor{black}{The isotropic single band analysis
\label{Sec - Hc2 analysis}}}

\begin{figure}
{\centering \resizebox*{0.48\textwidth}{!}{\includegraphics{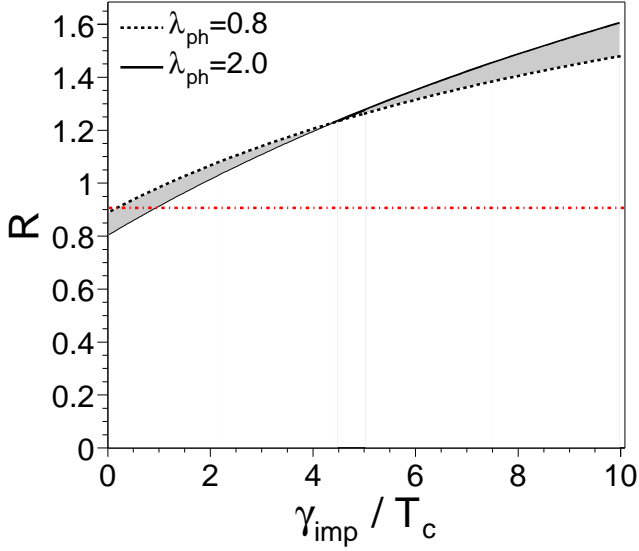}} \par}

\caption{\textcolor{black}{Parameter \protect\( R\protect \) vs. 
\protect\( \gamma _{\textrm{imp}}/T_{\textrm{c}}\protect \)
according to Eq. (\ref{Eq - R-check}) in the range of electron-phonon
coupling constants \protect\( 0.8\leq \lambda _{\textrm{ph}}\leq 2.0
\protect \).
Horizontal dash-dotted line: Experimental result for \protect\( 
\textrm{MgC}_{1.6}\textrm{Ni}_{3}\protect \)
derived from Eq. (\ref{Eq - R-checkexp}).}\label{Fig - R-check} }
\end{figure}

\textcolor{black}{In the following section the electron-phonon coupling
strength \( \lambda _{\textrm{ph}} \) is extracted from a simultaneous
analysis of the upper critical field and the penetration depth in
terms of the unknown impurity scattering rate \( \gamma _{
\textrm{imp}}\left[ \textrm{K}\right]  \).
Since the specific heat measurements do not clearly characterize \( 
\textrm{MgCNi}_{3} \)
as a one- or multi-band superconductor, the analysis starts within
an ISB model. Within this model the upper critical field \( H_{
\textrm{c}2}(0) \)
is given by\cite{shulga02}\begin{equation}
\label{Eq - ISBimpHc2}
H_{\textrm{c}2}(0)\left[ \textrm{Tesla}\right] =H^{\textrm{cl}}_{
\textrm{c}2}(0)\left[ 1+\frac{0.13\gamma _{\textrm{imp}}\left[ 
\textrm{K}\right] }{T_{\textrm{c}}(1+\lambda _{\textrm{ph}})}\right] ,
\end{equation}
where\begin{equation}
\label{Eq - ISBHc2}
H^{\textrm{cl}}_{\textrm{c}2}(0)\left[ \textrm{Tesla}\right] =0.0237
\frac{\left( 1+\lambda _{\textrm{ph}}\right) ^{2.2}T_{\textrm{c}}^{2}
\left[ \textrm{K}\right] }{v_{\textrm{F}}^{2}\left[ 10^{5}\frac{
\textrm{m}}{\textrm{s}}\right] },
\end{equation}
and \( \gamma _{\textrm{imp}}=v_{\textrm{F}}/l_{\textrm{imp}} \)
is the scattering rate which determines the intrinsic resistivity
(\( l_{\textrm{imp}} \) denotes the corresponding mean free path).
The London penetration depth including the unknown impurity scattering
rate \( \gamma _{\textrm{imp}} \) is given by an approximative formula
\begin{eqnarray}
\lambda _{\textrm{L}}(0) & \approx  & \widetilde{\lambda }_{
\textrm{L}}(0)\sqrt{\left( 1+\lambda _{\textrm{ph}}\right) 
\left( 1+0.7\frac{\gamma _{\textrm{imp}}}{2\Delta _{\textrm{exp}}}
\right) }\label{Eq - penetration_depth2a} \\
 & \qquad \equiv  & \widetilde{\lambda }_{\textrm{L}}(0)\sqrt{
\left( 1+\lambda _{\textrm{ph}}\right) \left( 1+0.7\frac{\gamma _{
\textrm{imp}}}{T_{\textrm{c}}}\frac{T_{\textrm{c}}}{2\Delta _{
\textrm{exp}}}\right) },\nonumber \label{Eq - penetration_depth2b} 
\end{eqnarray}
valid for \( \lambda _{\textrm{ph}}<2.5 \) (see App. \ref{appendix_A}
for the exact numerical expression), with the bare clean limit London
penetration depth\begin{equation}
\label{Eq - penetration_depth3}
\widetilde{\lambda }_{\textrm{L}}(0)=\frac{c}{\omega _{\textrm{pl}}}
\approx \frac{197.3\textrm{ nm}}{\omega _{\textrm{pl}}\textrm{ }
\left[ \textrm{eV}\right] }.
\end{equation}
Using Eqs. (\ref{Eq - ISBimpHc2}) and (\ref{Eq - penetration_depth2a}),
\( \gamma ^{\star }_{\textrm{N}}=\gamma _{0}(1+\lambda _{\textrm{ph}}) \)
and the experimentally determined quantities from Tab. 
\ref{Tab - main parameters},
we now will check the applicability of the ISB model. For this aim
we consider the ratio\begin{equation}
\label{Eq - R-checkexp}
R=\frac{6.77\times 10^{-6}\cdot \gamma ^{\star }_{\textrm{N}}\left[ 
\frac{\textrm{mJ}}{\textrm{molK}^{2}}\right] \cdot \lambda ^{2}_{
\textrm{L}}(0)\left[ \textrm{nm}^{2}\right] \cdot T^{2}_{
\textrm{c}}\left[ \textrm{K}^{2}\right] }{H_{\textrm{c}2}(0)
\left[ \textrm{Tesla}\right] \cdot V\left[ \textrm {\AA }^{3}\right] } 
\end{equation}
which includes the values of six experimentally readily accessible
quantities: the Sommerfeld coefficient \(  \gamma  ^{\star  }_{\textrm{N}} \),
\( H_{\textrm{c}2}(0) \), \( T_{\textrm{c}} \), \( \lambda _{\textrm{L}}(0) \)
and the volume of the unit cell. The dependence of \( R \) on the
parameter \( \gamma _{\textrm{imp}}/T_{\textrm{c}} \) can be expressed
as\begin{equation}
\label{Eq - R-check}
R=\frac{1+0.35\gamma _{\textrm{imp}}/\Delta (0)}{\left( 1+\lambda _{
\textrm{ph}}\right) ^{0.2}\left\{ 1+0.13\gamma _{\textrm{imp}}/\left[ T_{
\textrm{c}}\left( 1+\lambda _{\textrm{ph}}\right) \right] \right\} }
\end{equation}
In Fig. \ref{Fig - R-check}, the theoretical \( R(\gamma _{
\textrm{imp}}/T_{\textrm{c}}) \)
curves obtained from Eq. (\ref{Eq - R-check}) for several \( 
\lambda _{\textrm{ph}} \)
values are compared with the value of \( R \) derived from our experimental
data which is represented in Fig. \ref{Fig - R-check} as horizontal
line. Crossing points between the theoretical prediction and the experimental
result, which confirm the applicability of the ISB, are found for
\( \lambda _{\textrm{ph}}\geq 0.8 \) at low scattering rates. Even
in the case of higher electron-phonon coupling constants of \( \lambda _{
\textrm{ph}}\approx 2 \),
a clean limit scenario with \( \gamma _{\textrm{imp}}/T_{\textrm{c}}
\approx 1 \)
is favored within the ISB analysis. Dirty limit (with weak or medium
coupling) as proposed in Ref. \onlinecite{lin03} can be excluded from
the \( R \)-check in Fig. \ref{Fig - R-check}. }

\textcolor{black}{It is noteworthy that the proposed \( R \)-check
is much more convenient than the similar \( Q \)-check, proposed
recently by two of the present authors,\cite{shulga02} since the
dependence on \( \lambda _{\textrm{ph}} \) is considerably weaker
for \( R \) and, which is more important, \( R \) does not depend
on the band structure calculation. Thus comparing the results derived
above with the expectations from these calculations, additional information
on the nature of superconductivity in \( \textrm{MgCNi}_{3} \), may
be extracted. From Eqs. (\ref{Eq - ISBimpHc2}) and (\ref{Eq - ISBHc2}),
the effective Fermi velocity (in \( 10^{7}\textrm{ m}/\textrm{s} \))
\begin{eqnarray*}
v_{\textrm{F}} & = & 0.154\left( 1+\lambda _{\textrm{ph}}\right) ^{1.1}T_{
\textrm{c}}\left[ \textrm{K}\right] \times \nonumber \\
 &  & \times \sqrt{\frac{1+0.13\gamma _{\textrm{imp}}\left[ \textrm{K}
\right] /\left[ T_{\textrm{c}}\left[ \textrm{K}\right] \left( 1+\lambda _{
\textrm{ph}}\right) \right] }{H_{\textrm{c}2}(0)\left[ \textrm{Tesla}
\right] }}\label{Eq - vF-effective} 
\end{eqnarray*}
is obtained. Using the very weak scattering rates \( \gamma _{
\textrm{imp}}/T_{\textrm{c}}\leq 1 \)
derived above and the experimental values \( H_{\textrm{c}2}(0)=11
\textrm{ T} \)
and \( T_{\textrm{c}}=6.8\textrm{ K} \), one estimates from 
Eq. (\ref{Eq - vF-effective})
\( v_{\textrm{F}}\approx (0.60\ldots 1.08)\times 10^{7}\textrm{ cm}/
\textrm{s} \)
for electron-phonon coupling constants in the range of \( 0.8\leq 
\lambda _{\textrm{ph}}\leq 2.0 \).
Comparing this result with our band structure calculations (see Sec.
\ref{Sec 2}), one realizes consistence with the averaged \( v_{
\textrm{hc}2,\textrm{h}}=1.07\times 10^{7}\textrm{ m}/\textrm{s} \)
from the two hole Fermi surface sheets (Sec. \ref{Sec 2}) for strong
electron-phonon coupling of \( \lambda _{\textrm{ph}}\approx 2.0 \).
Thus, the relatively high value of the upper critical field of \( H_{
\textrm{c}2}(0)=11\textrm{ T} \)
can be attributed to strong electron-phonon coupling for the hole
subsystem. The second electron band plays a minor role for 
\( H_{\textrm{c}2}(0) \)
due to its much faster Fermi velocities and the much lower partial
density of states.}

\textcolor{black}{Having adopted the dominant hole picture, we also
can start from the band structure results, using the Fermi velocity
\( v_{\textrm{F},\textrm{h}} \) and the plasma frequency \( 
\omega _{\textrm{pl},\textrm{h}} \)
of the hole band. Then we have to find consistent values of \( 
\lambda _{\textrm{ph}} \)
and \( \gamma _{\textrm{imp}} \), which describe the \( H_{\textrm{c}2}(0) \)
and \( \lambda _{\textrm{L}}(0) \) data.}

\begin{figure}
{\centering \resizebox*{0.48\textwidth}{!}{\includegraphics{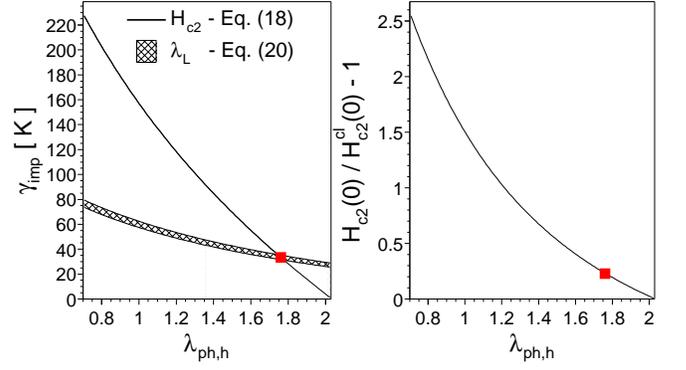}} \par}

\caption{\textcolor{black}{Left panel: Correlation between impurity scattering
rate \protect\( \gamma _{\textrm{imp}}\protect \) and electron-phonon
coupling constant \protect\( \lambda _{\textrm{ph},\textrm{h}}\protect \)
derived from Eq. (\ref{Eq - ISBimpHc2}) and 
Eq. (\ref{Eq - penetration_depth2a})
using \protect\( H_{\textrm{c}2}(0)=11\textrm{ T}\protect \), 
\protect\( T_{\textrm{c}}=6.8\textrm{ K}\protect \),
\protect\( v_{\textrm{F}}=1.07\times 10^{7}\textrm{ m}/\textrm{s}\protect \),
\protect\( \lambda _{\textrm{L}}(0)=237\textrm{ nm}\protect \), 
\protect\( 2\Delta (0)\approx 2.2\textrm{ meV}\protect \).
The point of intersection of both curves marked by a filled square
points to an electron-phonon coupling constant of \protect\( 
\lambda _{\textrm{ph},\textrm{h}}=1.74\ldots 1.78\protect \)
in the investigated \protect\( \textrm{MgC}_{1.6}\textrm{Ni}_{3}\protect \)
sample. Right panel: Ratio \protect\( \left[ \left( H_{
\textrm{c}2}(0)/H^{\textrm{cl}}_{\textrm{c}2}(0)\right) -1\right] \protect \)
plotted against \protect\( \lambda _{\textrm{ph},\textrm{h}}\protect \).
The filled square again corresponds to \protect\( \lambda _{\textrm{ph},
\textrm{h}}=1.74\ldots 1.78\protect \).
From \protect\( H_{\textrm{c}2}(0)=11\textrm{ T}\protect \) one estimates
\protect\( H^{\textrm{cl}}_{\textrm{c}2}(0)\approx (8.79\ldots 9.07)
\textrm{ T}\protect \)
for the upper critical field in the clean limit. 
\label{Fig - impurityscattering-coupling}}}
\end{figure}

\textcolor{black}{From the plasma frequency of band \( 1 \), \( 
\hbar \omega _{\textrm{pl},1}=1.89\ldots 1.94\textrm{ eV} \)
(see Sec. \ref{Sec 2}), we get \( \widetilde{\lambda }_{
\textrm{L}}(0)=\left( 101.7\ldots 104.4\right) \textrm{ nm} \),
using Eq. (\ref{Eq - penetration_depth3}). With the empirical values
of \( \lambda _{\textrm{L}}(0)=237\textrm{ nm} \) and \( 2\Delta (0)
\approx 2\Delta _{\textrm{exp}}=2.2\textrm{ meV}\doteq 25.5\textrm{ K} \)
for the superconducting gap (see Sec. \ref{Sec - sl state analysis}),
Eq. (\ref{Eq - penetration_depth2a}) depends only on \( \gamma _{
\textrm{imp}} \)
and \( \lambda _{\textrm{ph},\textrm{h}} \) (of the hole band). The
same applies to Eq. (\ref{Eq - ISBimpHc2}), using the experimental
values \( H_{\textrm{c}2}(0)=11\textrm{ T} \), \( T_{\textrm{c}}=6.8
\textrm{ K} \)
and the calculated average Fermi velocity of the hole band, \( v_{
\textrm{hc}2,\textrm{h}}=1.07\times 10^{7}\textrm{ m}/\textrm{s} \).
The correlation between \( \gamma _{\textrm{imp}} \) and \( \lambda _{
\textrm{ph},\textrm{h}} \),
emerging from these two equations, is shown in the left panel of Fig.
\ref{Fig - impurityscattering-coupling}. The intersection of both
graphs gives \( \lambda _{\textrm{ph},\textrm{h}}=1.74\ldots 1.78 \)
and \( \gamma _{\textrm{imp}}=(31.0\ldots 36.0)\textrm{ K} \). Thus,
we arrive at a higher, more realistic scattering rate compared with
\( \gamma _{\textrm{imp}}\approx T_{\textrm{c}} \) obtained from
the \( R \)-check in Fig. \ref{Fig - R-check}. The corresponding
ratio \( \left[ \left( H_{\textrm{c}2}(0)/H^{\textrm{cl}}_{
\textrm{c}2}(0)\right) -1\right]  \),
giving the deviation of \( H_{\textrm{c}2}(0) \) from the clean limit
value \( H^{\textrm{cl}}_{\textrm{c}2}(0) \), is plotted in the right
panel of Fig. \ref{Fig - impurityscattering-coupling}. One gets \( H^{
\textrm{cl}}_{\textrm{c}2}(0)\approx (8.79\ldots 9.07)\textrm{ T} \). }

\textcolor{black}{To summarize this part, already in the simplest
possible approach two general properties of \( \textrm{MgCNi}_{3} \)
are derived:}

\begin{enumerate}
\item \textcolor{black}{strong electron-phonon coupling and}
\item \textcolor{black}{intrinsic clean limit at least for the hole subsystem.}
\end{enumerate}
\textcolor{black}{Nevertheless, it should be noted that recent preliminary
measurements of the penetration depth by \citeauthor{lin03},\cite{lin03}
resulting in \( \lambda _{\textrm{L}}(0)=(128\ldots 180)\textrm{ nm} \)
are not compatible with the presented effective single band analysis
(see also Ref. \onlinecite{lin}). Especially the \( R \)-check 
(Eqs. (\ref{Eq - R-checkexp}) and (\ref{Eq - R-check})) results in unphysical
solutions (\( \lambda _{\textrm{ph}}=30 \) as a lower limit), using
the values presented in Ref. \onlinecite{lin03} and \onlinecite{lin}
(see as well Tab. \ref{Tab - main parameters}). The consequences,
if these measurements could be verified, remain unclear.}

\subsection{\textcolor{black}{Strong coupling and enhanced depairing}
\textcolor{blue}{\label{Section - sfph-discussion}}}

\begin{table}

\caption{Characteristic phonon frequency and coupling parameters derived by
analyzing the experimental data of the present \protect\( 
\textrm{MgC}_{1.6}\textrm{Ni}_{3}\protect \)
sample.\label{Tab - coupling-results} }

\begin{ruledtabular}

{\centering \begin{tabular}{ccdll}
&
&
\multicolumn{2}{c}{c\( _{\textrm{p}} \) analysis}&
\multicolumn{1}{c}{\( H_{\textrm{c}2} \) analysis}\\
&
\multicolumn{1}{p{1.4cm}}{}&
\multicolumn{1}{l}{\parbox[t]{2cm}{normal state \\
(Sec. \ref{Sec - normal state cp analysis})}}&
\multicolumn{1}{l}{\parbox[t]{2cm}{ sl state\\
(Sec. \ref{Sec - sl state analysis})}}&
\multicolumn{1}{l}{\parbox[t]{2cm}{~ \\
(Sec. \ref{Sec - Hc2 analysis})}}\\
\cline{4-4} \cline{5-5} 
\hline 
\( \omega _{\textrm{ln}} \)&
\multicolumn{1}{c}{\( \left[ \textrm{K}\right]  \)}&
143&
\multicolumn{1}{c}{\( 88\ldots 134 \)}&
\\
\( \lambda _{\textrm{ph}} \)&
&
1.85&
\multicolumn{1}{c}{\( 1.9\ldots 2.3 \)}&
\( 1.74\ldots 1.78 \)\footnotemark[1]\\
\( \lambda _{\textrm{sf}} \)&
&
0.43&
&
\\
\end{tabular}\par}

\end{ruledtabular}

\footnotetext[1]{Limited to band {}``1''.}
\end{table}
Several results of our analysis of the experimental data are summarized
in Table \ref{Tab - coupling-results}. The comparison of the estimated
\( \lambda _{\textrm{ph}} \) values clearly points to strong electron-phonon
coupling. Nevertheless, the strong coupling scenario realized in \( 
\textrm{MgCNi}_{3} \)
has been questioned.\cite{he01,lin03} The strong electron-phonon
coupling found for \( \textrm{MgCNi}_{3} \) requires a sizable depairing
contribution to explain the low \( T_{\textrm{c}} \) value, otherwise
at least a twice as large \( T_{\textrm{c}} \) would be expected.
It is illustrative to compare different approaches for the calculation
of \( T_{\textrm{c}} \) to analyze the electron-phonon coupling strength
under consideration of the low temperature upturn of the specific
heat in the normal state (see Sec. \ref{Sec - normal state cp analysis}). 

In a first approach usually the low temperature Debye approximation
is used to extract the Debye temperature which we did in 
Sec. \ref{Sec - cp results}
for comparison. Our result of 
\( \Theta _{\textrm{D}}^{\star }=292\textrm{ K} \)
is in agreement with previous measurements of \citeauthor{lin03},
\cite{lin03}
\citeauthor{mao03}\cite{mao03} and calculations of 
\citeauthor{Ignatov03}\cite{Ignatov03}
(It should be noted, that our specific heat data were corrected by
carbon contribution (see Fig. \ref{xray Diffraktogramm}), without
this correction we arrive at \( \Theta _{\textrm{D}}^{\star }=285
\textrm{ K} \)).
In this analysis the McM\textcolor{black}{illan formula\begin{equation}
\label{Eq - McMillanformulae}
T_{\textrm{c}}=\frac{\omega _{\textrm{D}}}{1.45}\exp \left[ -1.04
\frac{1+\lambda _{\textrm{ph}}}{\lambda _{\textrm{ph}}-\mu ^{\star }
\left( 1+0.62\lambda _{\textrm{ph}}\right) }\right] 
\end{equation}
is usually applie}d. This approximation is only useful for a special
phonon spectrum with \( \omega _{\textrm{ln}}/\omega _{\textrm{D}}
\approx 0.6 \).
In the case \textcolor{black}{of \( \textrm{MgCNi}_{3} \) we found
\( \approx 0.30\ldots 0.49 \) (cor}responding to \( \omega _{
\textrm{ln}}\approx 88\ldots 143\textrm{ K} \))
and the Allen-Dynes formula (Eq. (\ref{Eq - Allen-Dynes})) should
be applied instead. 
\begin{figure}
{\centering \resizebox*{0.48\textwidth}{!}{\includegraphics{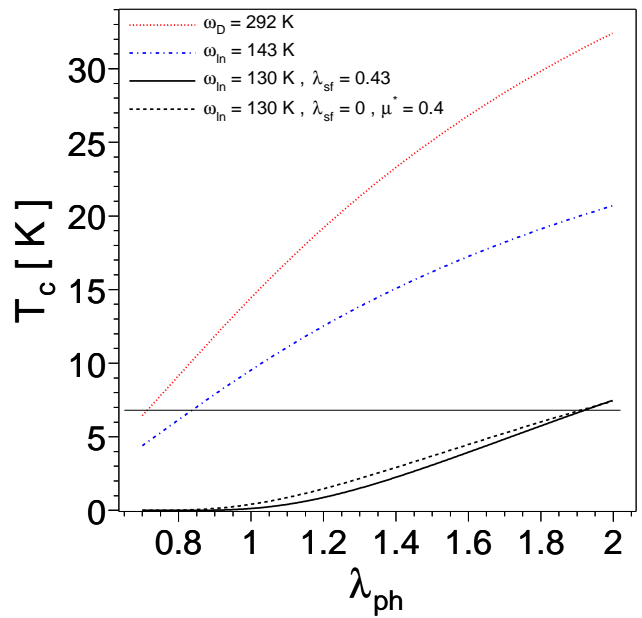}} \par}

\caption{Variation of \protect\( T_{\textrm{c}}\protect \) with \protect\( 
\lambda _{\textrm{ph}}\protect \)
with\textcolor{black}{out electron-paramagnon interaction and {}``normal''
Coulomb pseudopotential \protect\( \mu ^{\star }=0.13\protect \)
accordin}g to Eq. (\ref{Eq - McMillanformulae}) (dotted line) and
Eq. (\ref{Eq - Allen-Dynes}) (dash-dotted line) and with enhanced
pair-breaking contribution according to Eq. (\ref{Eq - Allen-Dynes-sf})
by \protect\( \mu ^{\star }=0.13\protect \) and \protect\( \lambda _{
\textrm{sf}}=0.43\protect \)
(solid line), respectively \protect\( \mu ^{\star }=0.4\protect \)
and \protect\( \lambda _{\textrm{sf}}=0\protect \) (dashed line).
The characteristic phonon frequencies are chosen from Sec. 
\ref{Sec - cp results}
(dotted line), Sec. \ref{Sec - normal state cp analysis} (dash-dotted
line), respectively Sec. \ref{Sec - sl state analysis} (solid and
dashed line). The points of intersection of the curves with the horizontal
line at \protect\( T_{\textrm{c}}=6.8\textrm{ K}\protect \) show
the electron-phonon coupling strengths \protect\( \lambda _{
\textrm{ph}}\protect \)
resulting in the different approaches.\label{Fig - mcmillan-allendynes}}
\end{figure}

\textcolor{black}{Fig. \ref{Fig - mcmillan-allendynes} compares both
equations using \( \Theta _{\textrm{D}}^{\star }\equiv \omega _{
\textrm{D}}=292\textrm{ K} \)
(dotted line) respectively \( \omega _{\textrm{ln}}=143\textrm{ K} \)
(dash-dotted line). In both cases the Coulomb pseudopotential was
fixed to \( \mu ^{\star }=0.13 \). Apart from the deviation between
Eq. (\ref{Eq - Allen-Dynes}) and Eq. (\ref{Eq - McMillanformulae})
due to the ratio \( \omega _{\textrm{ln}}/\omega _{\textrm{D}}\leq 0.49 \),
both equations seem to result in a moderate electron-phonon coupling
of \( \lambda _{\textrm{ph}}=0.67\ldots 0.82 \) if no additional
pair breaking effects} are considered.

\textcolor{black}{However, we remind the reader, that the experimental
and theoretical picture of \( \textrm{MgCNi}_{3} \) strongly indicates
strong electron-phonon coupling and a spin fluctuation contribution.
The solid line compared to the dash-dotted line in Fig. 
\ref{Fig - mcmillan-allendynes}
shows that the dependence of \( T_{\textrm{c}} \) on \( 
\lambda _{\textrm{ph}} \)
is strongly influenced by pair-breaking contributions such as the
presence of enhanced electron-paramagnon coupling \( 
\lambda _{\textrm{sf}}=0.43 \).
The same situation in the imaginable case of purely static pair-breaking,
expressed by \( \mu ^{\star }=0.4 \) is given by the dotted line.
A very similar result was reported by \citeauthor{Ignatov03}\cite{Ignatov03}
who proposed a phonon-softening scenario with \( T_{\textrm{c}}=8
\textrm{ K} \),
\( \omega _{\textrm{ln}}=120\textrm{ K} \), \( \lambda _{\textrm{ph}}=1.51 \)
and an enhanced \( \mu ^{\star }=0.33 \) due to spin fluctuations.
In any case the superconducting transition temperature is strongly
suppressed by pair-breaking contributions.}

\textcolor{black}{\( T_{\textrm{c}} \) of \( \textrm{MgCNi}_{3} \)
would rise up to \( \approx 20\textrm{ K} \), if one somehow could
suppress the electron-paramagnon interaction. In that case the 
electron-phonon
coupling would not be affected and the dash-dotted line in Fig. 
\ref{Fig - mcmillan-allendynes}
would become reality.}

\textcolor{black}{Within the phonon-softening scenario,\cite{Ignatov03}
which was recently observed in neutron-scattering measurements,\cite{heid03}
a part o}f the low temperature specific heat anomaly may be of phonon
origin (as stated in Sec. \ref{Sec - normal state cp analysis}).
In this picture the electron-paramagnon coupling would be reduced
with the possibility of a paramagnon shift to higher temperature
\textcolor{black}{s.
This is con}sistent with \( \omega _{\textrm{ln}}\approx 100\textrm{ K} \)
(lower limit of the result from Sec. \ref{Sec - sl state analysis})
and an electron-paramagnon coupling constant of \( \lambda _{
\textrm{sf}}\approx 0.25 \).
Using these numbers in Eq. (\ref{Eq - Allen-Dynes-sf}) the electron-phonon
coupling constant amounts \( \lambda _{\textrm{ph}}\approx 1.6 \).
\textcolor{black}{To find the composition of the phonon and paramagnon
contribution to the upturn, low temperature neutron-scattering measurements
should be performed. In this context we remind the reader, that spin
fluctuations are known to show a dependence on the applied magnetic
field, which indeed is seen in Fig. \ref{Bild - spezifische 
W=E4rme H=3D0 inset}. }

\subsection{\textcolor{black}{Multi-band effects beyond the standard 
approach\label{Sec D - multiband}}}

\textcolor{black}{Multi- (two-) band (and similar anisotropy) effects
for several physical properties in the superconducting state are in
principle well known for a long time,\cite{moskalenko73} especially
for weakly coupled superconductors in the clean limit. To the best
of our knowledge their interplay with disorder and strong coupling
effects is less systematically studied. In particular this is caused
by the increased number of input parameters and the necessity of a
large amount of numerical calculations.}

\textcolor{black}{The multi-band character in \( \textrm{MgCNi}_{3} \)
is not only supported by the band structure calculations, but also
by experimental findings, and for instance, reflected by the large
gap found in tunneling measurements\cite{mao03,shan03pc} and a smaller
one seen in NMR measurements.\cite{singer01}}

\textcolor{black}{Like in \( \textrm{MgB}_{2} \) the effect of interband
scattering is expected to be weak due to the presence of well disjoint
FSS. However, due to different contributions of the partial density
of states compared to the case of \( \textrm{MgB}_{2} \), the two-band
character of \( \textrm{MgCNi}_{3} \) is less pronounced. In Sec.
\ref{Sec - sl state analysis} the total electron-phonon coupling
constant averaged over all Fermi surface sheets, \( \lambda _{
\textrm{ph},\textrm{tot}} \)
was estimated by Eq. (\ref{Eq - Sommerfeld}) at \( \lambda _{
\textrm{ph},\textrm{tot}}\approx 1.9 \).
Considering the band structure calculation presented in Sec. \ref{Sec 2},
this value is to be distributed among the two contributing bands according
to\begin{equation}
\label{Eq - decomposing coupling}
\lambda _{\textrm{ph},\textrm{tot}}=\lambda _{\textrm{h}}\frac{N_{
\textrm{h}}(0)}{N(0)}+\lambda _{\textrm{el}}\frac{N_{\textrm{el}}(0)}{N(0)}.
\end{equation}
With \( \lambda _{\textrm{ph},\textrm{h}}=1.74\ldots 1.78 \) (see
Sec. \ref{Sec - Hc2 analysis}), the coupling in the second band amounts
\( \lambda _{\textrm{ph},\textrm{el}}=2.58\ldots 2.81 \). Obviously
this strong mass enhancement in both bands is not compatible with
the low value of \( T_{\textrm{c}}=6.8\textrm{ K} \). So, as in the
single band case a pair breaking contribution is needed. }

\begin{figure}
{\centering \resizebox*{0.48\textwidth}{!}{\includegraphics{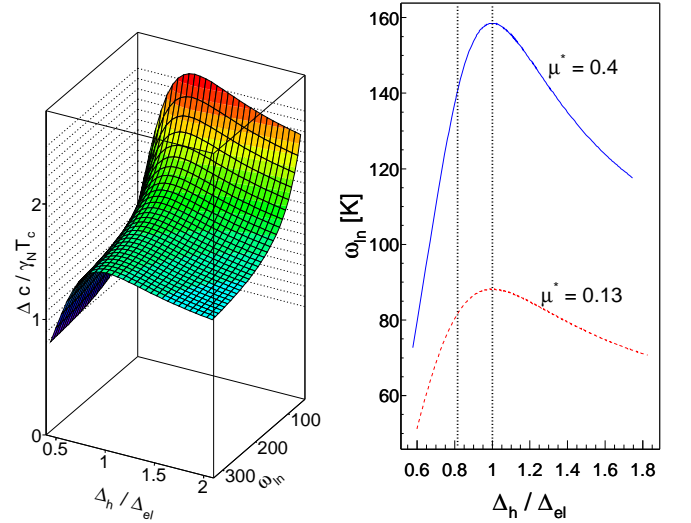}} \par}

\caption{Left panel: Dependence of the specific heat jump \protect\( 
\Delta \textrm{c}/\gamma _{\textrm{N}}T_{\textrm{c}}\protect \)
on the gap ratio \protect\( \Delta _{\textrm{h}}/\Delta _{\textrm{el}}
\protect \)
and the characteristic phonon frequency \protect\( \omega _{\textrm{ln}}
\protect \)
within the two-band description for \protect\( \mu ^{\star }=0.13\protect \)
(Eq. (\ref{Eq - twobandjump})). Right panel: Possible solutions for
\protect\( \Delta _{\textrm{h}}/\Delta _{\textrm{el}}\protect \)
and \protect\( \omega _{\textrm{ln}}\protect \) to reach the experimental
specific heat jump for the two cases \protect\( \mu ^{\star }=0.13\protect \)
and \protect\( \mu ^{\star }=0.4\protect \) (enhanced pair-breaking).
The two vertical lines mark the range of expected \protect\( \Delta _{
\textrm{h}}/\Delta _{\textrm{el}}\protect \)
values (see Sec. \ref{Sec - sl state analysis}).\label{Fig - jump-2band}}
\end{figure}

\textcolor{black}{Measurable quantities describing the superconducting
transition in the case of \( \textrm{MgCNi}_{3} \) (particularly
Eqs. (\ref{carbotte:alle})) are affected in opposite ways by strong
coupling effects from one side and two-band effects from the other
sides.} Here we will briefly show, how these different effects influence
the \textcolor{black}{specific heat jump \( \Delta \textrm{c}/\gamma _{
\textrm{N}}T_{\textrm{c}} \)}
according to Eq. (\ref{carbotte1}). Considering these effects,
\begin{equation}
\label{Eq - twobandjump}
\frac{\Delta \textrm{c}}{\gamma T_{\textrm{c}}}=1.43\textrm{F}(\mu ^{
\star })\frac{\textrm{B}_{1}\left( \frac{\omega _{\textrm{ln},
\textrm{h}}}{T_{\textrm{c}}}\right) \left( 1+\eta v/z^{2}\right) ^{2}}{
\left( 1+v\right) \left( 1+\eta v/z^{4}\right) }
\end{equation}
is derived, where\begin{equation}
\label{Eq - nu}
v=\frac{\left( 1+\lambda _{\textrm{ph},\textrm{el}}\right) N_{
\textrm{el}}}{\left( 1+\lambda _{\textrm{ph},\textrm{h}}\right) N_{
\textrm{h}}}
\end{equation}
contains electron-phonon coupling and multi-band corrections and
\begin{equation}
\label{Eq - z}
z=\frac{\Delta _{\textrm{h}}}{\Delta _{\textrm{el}}}
\end{equation}
denotes the gap-ratio. The function \( \textrm{F}(\mu ^{\star }) \),
given by\[
\textrm{F}(\mu ^{\star })=\frac{1.15\left[ 1+0.156\tanh \left( 5\mu ^{
\star }\right) \right] }{1+0.156\tanh \left( 0.5\right) }\]
has been obtained by analyzing numerical data derived by 
\citeauthor{carbotte90}\cite{carbotte90}
under consideration of enhanced pair-breaking in terms of the Coulomb
pseudopotential \( \mu ^{\star } \) (valid up to \( \mu ^{\star }
\approx 0.4 \)).
The different electron-phonon coupling constants in both bands may
involve different characteristic phonon frequencies \( \omega _{
\textrm{ln}} \).
This situation is formally taken into account by the parameter 
\( \eta  \),
given by\[
\eta =\textrm{B}_{1}\left( \frac{\omega _{\textrm{ln},
\textrm{el}}}{T_{\textrm{c}}}\right) /\textrm{B}_{1}\left( 
\frac{\omega _{\textrm{ln},\textrm{h}}}{T_{\textrm{c}}}\right) .\]
The general result, depending on the gap ratio \( z \) and the characteristic
phonon frequency \( \omega _{\textrm{ln}} \) (for simplification
\( \omega _{\textrm{ln},\textrm{el}}=\omega _{\textrm{ln},\textrm{h}} \)
is assumed) is shown in the left panel of Fig. \ref{Fig - jump-2band}
(with \textcolor{black}{\( \lambda _{\textrm{ph},\textrm{h}}=1.76 \),}
\textcolor{black}{\( \lambda _{\textrm{ph},\textrm{el}}=2.7 \)} and
\( \textrm{F}(\mu ^{\star })=1 \)). The opposite effect of strong
coupling from one side and two-band behavior from the other side is
clearly seen. The right panel shows possible solutions for \( z \)
and \( \omega _{\textrm{ln}} \) to reach the experimental determined
jump \textcolor{black}{\( \Delta \textrm{c}/\gamma _{\textrm{N}}T_{
\textrm{c}}=2.09 \).
Considering the results from the normal state specific heat analysis
(\( \omega _{\textrm{ln}}\approx 143\textrm{ K} \)) and the superconducting
specific heat analysis (}\( 1\leq z\leq 0.8 \)\textcolor{black}{),
best consistency is obtained with} \( \mu ^{\star }\approx 0.4 \)
in full agreement with the pair-breaking scenario. 

\begin{figure}
{\centering \resizebox*{0.48\textwidth}{!}{\includegraphics{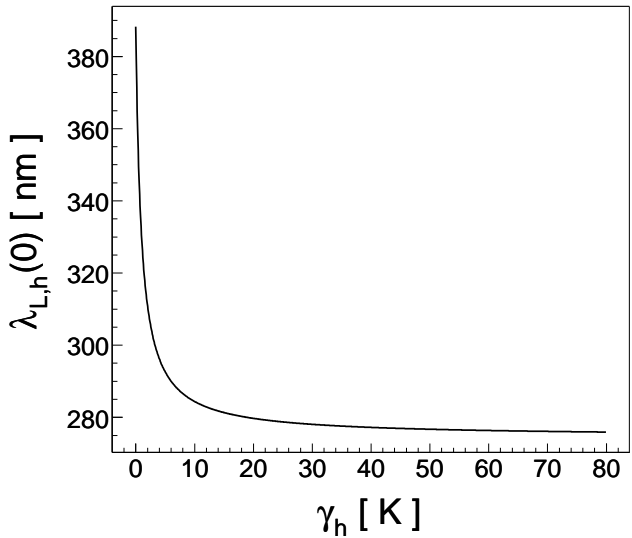}} \par}

\caption{Penetration depth at zero temperature derived from 
Eq. (\ref{Eq - penetrationdepth-2band})
vs. scattering rate in the hole band using \protect\( \gamma _{
\textrm{el}}/\gamma _{\textrm{h}}\approx 4.81\protect \)
(see Eq. (\ref{Eq - scatteringrateratio})).
\label{Fig - calculated penetration depth}}
\end{figure}

Now, the influence of two-band corrections on the penetration depth
\( \lambda _{\textrm{L}}(0) \) should be checked, since the experimental
value of \( \lambda _{\textrm{L}}(0)=237\textrm{ nm} \) was ascribed
to the hole band in Sec. \ref{Sec - Hc2 analysis}. We start with
the inverse squared total penetration depth of a two-band superconductor.
It is given as a sum of the two corresponding contributions from each
band:\[
\lambda _{\textrm{tot}}^{-2}(0)=f_{\textrm{h}}\lambda _{\textrm{h}}^{-2}(0)+f_{
\textrm{el}}\lambda _{\textrm{el}}^{-2}(0),\]
with the gap ratio dependent weighting factors \( f_{\textrm{i}}(z,\lambda _{
\textrm{i}},N_{\textrm{i}}) \),
where \( \textrm{i}=\textrm{h},\textrm{ el} \) and \( f_{\textrm{i}}
\equiv 1 \)
in the case of equal gaps (i.e. \( z=1 \)). The corresponding specific
plasma frequencies, coupling constants, and gaps do enter each term
(see Eqs. (\ref{Eq - penetration_depth2a},\ref{Eq - penetration_depth3})).
Then the total penetration depth can be rewritten as{\small \hfill{}
\begin{eqnarray}
\lambda _{\textrm{L},\textrm{h}}(0) & = & \lambda _{\textrm{L}}(0)
\times \nonumber \\
 &  & \times \sqrt{\textrm{L}\left[ 1+\frac{\Delta _{\textrm{el}}^{2}
\omega ^{2}_{\textrm{pl},\textrm{el}}(1+\lambda _{\textrm{h}})(1+\frac{0.7
\gamma _{\textrm{h}}}{2\Delta _{\textrm{h}}})}{\Delta _{\textrm{h}}^{2}
\omega ^{2}_{\textrm{pl},\textrm{h}}(1+\lambda _{\textrm{el}})(1+\frac{0.7
\gamma _{\textrm{el}}}{2\Delta _{\textrm{el}}})}\right] }
\label{Eq - penetrationdepth-2band} 
\end{eqnarray}
}with\[
\textrm{L}=\frac{1+v/z^{2}}{1+v/z^{4}}.\]
With the above determined values, we get \( \textrm{L}\approx 0.88 \)
(with \( \nu  \) and \( z \) according to Eqs. (\ref{Eq - nu},\ref{Eq - z})).
Fig. \ref{Fig - calculated penetration depth} shows the contribution
of the hole band to \( \lambda _{\textrm{L}}(0) \) for the LDA calculation,
using \( \gamma _{\textrm{el}}/\gamma _{\textrm{h}}\approx 4.81 \)
(see Eq. (\ref{Eq - scatteringrateratio})). It is seen, that \( \lambda _{
\textrm{L},\textrm{h}}(0) \)
exceeds the experimental value of \( \lambda _{\textrm{L}}(0)=237
\textrm{ nm} \)
b\textcolor{black}{y no more than \( \approx 18\textrm{ }\% \) (in
the case of \( \gamma _{\textrm{h}}=31.0\ldots 36.0\textrm{ K} \)
-- see S}ec. \ref{Sec - Hc2 analysis}), indicating only small influence
of the electron band on \( \lambda _{\textrm{L}}(0) \). Nevertheless
its influence is not as small as in the case of \( H_{\textrm{c}2}(0) \)
and thus, we estimate an error of about \( \approx 10\textrm{ }\% \)
for the electron-phonon coupling constant of the hole band, determined
in Sec. \ref{Sec - Hc2 analysis}.

\section{Conclusions}

\textcolor{black}{Our analysis of \( \textrm{MgCNi}_{3} \) revealed
a highly interesting interplay of different, on first glance unexpected
adversed physical features or tendencies all present within one material
causing a rather complex general behavior. This novel superconductor
has been interpreted so far as standard \( s \)-wave BCS superconductor
or as unconventional superconductor with strong or medium electron-phonon
coupling. Last but not least, considerable pair-breaking contribution
due to spin fluctuations and / or Coulomb repulsion have been suggested
from theory and experiment.}

\textcolor{black}{The present analysis is the first approach to reconcile
the unusual experimental findings within a unified physical picture.
It reveals strong electron-phonon coupling combined with medium 
electron-paramagnon
coupling. Strong electron-phonon coupling was derived from specific
heat data in the normal and superconducting state independently. An
unusual upturn of the specific heat in the normal state observed at
low temperatures can be attributed to spin fluctuations and / or a
softening of low-frequency phonons. To specify the contribution of
the belonging electron-boson interactions to the low temperature specific
heat anomaly, low-temperature neutron measurements are necessary in
order to investigate the evolution of the lattice excitations, which
may even be modified by the transition from the normal to the superconducting
state.}

\textcolor{black}{The electronic specific heat data show an exponential
temperature dependence at low temperatures which is a strong indication
for \( s \)-wave superconductivity in \( \textrm{MgCNi}_{3} \).
It was shown that a contribution of a second band could not be excluded
but even complies with recent tunneling measurement results. The multi-band
character of \( \textrm{MgCNi}_{3} \) is proved by band structure
calculations. However, with respect to superconductivity the two-band
character of \( \textrm{MgCNi}_{3} \) is much less pronounced than
in the model compound \( \textrm{MgB}_{2} \). That is due to the
predominance of a hole band with a large density of states in 
\( \textrm{MgCNi}_{3} \),
whereas in \( \textrm{MgB}_{2} \) the densities of states of both
bands are comparable. Therefore, several properties such as the specific
heat or the upper critical field can be described to first approximation
reasonably well within an effective single band model. Nevertheless,
other properties such as the Hall conductivity and the thermopower
require a multi-band description, i.e. at least one effective electron
and one effective hole band (see App. \ref{appendix_B}). Previous
theoretical analyses based on single-band models could describe only
few physical properties. As a consequence of the oversimplified approaches
they blamed the local density approximation to fail seriously. This
is in sharp contrast to our analysis of the upper critical field yielding
an effective Fermi velocity agreeing well with the LDA hole band prediction.
Our proposed effective strong coupling two-band approach explains
the complex behavior observed for \( \textrm{MgCNi}_{3} \) and is
expected to hold for other still not examined physical properties.}

\textcolor{black}{The highly interesting interplay of strong electron-phonon
coupling on multiple Fermi surface sheets, softening of lattice excitations,
the strong energy dependence of the density of states near the Fermi
energy of one band (van Hove singularity), and paramagnons or strong
Coulomb repulsion for a realistic, anisotropic multi-band electronic
structure with nesting features in this compound highly motivates
further experimental studies. Investigating the influence of impurities
or slight stoichiometry-deviations on the electronic and bosonic properties
would be as helpful as making of purer samples and single crystals
to perform quantum oscillation studies like de Haas van Alphen measurements.}

\textcolor{black}{Deepened theoretical studies are needed to clarify
remaining quantitative details and to extend the present-day strong
coupling Eliashberg theory with all peculiarities of \( 
\textrm{MgCNi}_{3} \). }

\begin{appendix}

\section{\label{appendix_A}Penetration depth : strong coupling and impurity
scattering}

\begin{figure}
{\centering \resizebox*{0.48\textwidth}{!}{\includegraphics{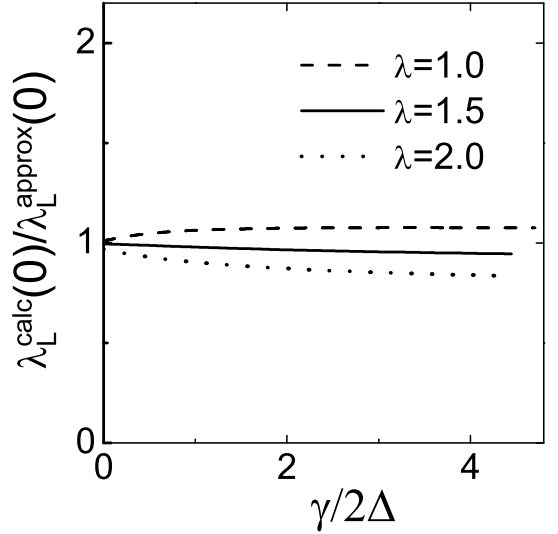}} \par}

\caption{Results of strong coupling calculations for the penetration depth
at zero temperature (Eq. (\ref{A1})) for several electron-phonon
coupling constants \protect\( \lambda \protect \) vs. impurity scattering
rate \protect\( \gamma _{\textrm{imp}}\protect \) (in units of the
gap \protect\( \Delta _{\textrm{exp}}=1.1\textrm{ meV}\protect \)
as derived from Sec. \ref{Sec - sl state analysis}) in comparison
with the approximate expression provided by 
Eq. (\ref{Eq - penetration_depth2a}).\label{penetrationEliashberg}}
\end{figure}

We present a simple semi-analytic expression for the penetration depth
at \( T=0\textrm{K} \) for type-II superconductors valid in the London
limit. Thereby strong coupling and impurity scattering effects are
treated on equal footing within standard Eliashberg theory. In calculating
\( \lambda _{\textrm{L}}(0) \) we start from an expression proposed
first by \citeauthor{nam67}\cite{nam67} and later on frequently used
in the literature\cite{mar90,adr95,gol02}\begin{equation}
\label{A1}
\lambda ^{-2}_{\textrm{L}}(0)=\frac{\pi T\omega _{\textrm{pl}}^{2}}{
\textrm{c}^{2}}\Sigma _{n=1}^{\infty }\frac{\Delta ^{2}\left( 
\textrm{i}\omega _{n}\right) }{Z\left( \textrm{i}\omega _{n}\right) 
\left[ \omega ^{2}_{n}+\Delta ^{2}\left( \textrm{i}\omega _{n}\right) 
\right] ^{3/2}},
\end{equation}
where \( \textrm{i}\omega _{n}=\textrm{i}\pi (2n-1)T \), 
\( n=0,\pm 1,\pm 2,\ldots  \)
are the Matsubara frequencies and \( Z(\textrm{i}\omega _{n}) \)
as well as \( \Delta (\textrm{i}\omega _{n}) \) denote the renormalization
factor and the gaps, respectively. The result of our numerical calculations
of Eq. (\ref{A1}) compared with the approximation given by 
Eq. (\ref{Eq - penetration_depth2a})
is shown in Fig. \ref{penetrationEliashberg}.

One realizes only small deviations not exceeding \( 8 \) to \( 10
\textrm{ }\% \)
which is sufficient for our qualitative estimate of large mean free
paths \( l_{\textrm{imp}} \) compared with the coherence length 
\( \xi _{\textrm{GL}}(0) \).

\section{\label{appendix_B}Two-band influence on other physical quantities}

With the help of the two-band scenario even difficulties found explaining
Hall data can be overcome.\cite{li01,young03} Within the two-band
model (Sec. \ref{Sec 2}) the Hall constant is defined as\[
R_{\textrm{H}}=\frac{R_{\textrm{H},\textrm{el}}\sigma ^{2}_{
\textrm{el}}+R_{\textrm{H},\textrm{h}}\sigma _{\textrm{h}}^{2}}{
\left( \sigma _{\textrm{el}}+\sigma _{\textrm{h}}\right) ^{2}},\]
using \( R_{\textrm{H},\textrm{el}}=-R_{\textrm{H},\textrm{h}} \)
with \( n_{\textrm{h}}=n_{\textrm{el}} \) (due to the even number
of electrons per unit cell) and the ratio of the hole and electron
conductivities\[
x=\frac{\sigma _{\textrm{h}}}{\sigma _{\textrm{el}}}\approx \frac{N_{
\textrm{h}}v^{2}_{\textrm{F},\textrm{h}}\gamma _{\textrm{el}}}{N_{
\textrm{el}}v^{2}_{\textrm{F},\textrm{el}}\gamma _{\textrm{h}}},\]
where \( \gamma _{\textrm{h}} \) and \( \gamma _{\textrm{el}} \)
are the corresponding scattering rates, we get\begin{equation}
\label{Eq - Hall2}
R_{\textrm{H}}=R_{\textrm{H},\textrm{el}}\frac{1-x}{1+x},
\end{equation}
with \( R_{\textrm{H},\textrm{el}}=-1/\left( n_{\textrm{el}}ec\right)  \).
The number of charges per unit cell from LDA-FPLO calculations (see
Sec. \ref{Sec 2}) amounts \( n=0.285 \) (comparable to \( n=0.303 \)
of Ref. \onlinecite{kumary02}). The resulting theoretical charge carrier
density of \( n_{\textrm{el}}\approx 5.2\times 10^{21}/\textrm{cm}^{3} \)
and the calculated conductivity ratio of \( x=1.403 \) (see Sec.
\ref{Sec 2}) should now be compared with measurements in terms of
the effective Hall constant given by Eq. (\ref{Eq - Hall2}).

Since so far reported samples are widely spread in terms of their
residual resistivities, grain boundary effects, affecting the Hall
conductivity should be taken into account. This can be done in a first
approximation by analyzing the mean free path (Eq. (\ref{Formel - 
mittlere freie Weglaenge 2})),
considering the hole subsystem\[
l_{\textrm{imp},\textrm{h}}=4.9\times 10^{2}\frac{v_{\textrm{F},
\textrm{h}}\left[ 10^{7}\textrm{ cm}/\textrm{s}\right] }{\left( 
\omega _{\textrm{pl},\textrm{h}}\textrm{ }\left[ \textrm{eV}
\right] \right) ^{2}\rho _{0}\left( 1+\frac{1}{x}\right) 
\textrm{ }\left[ \mu \Omega \textrm{cm}\right] }.\]

Similar considerations should be applied to the analysis of the thermopower,
where also an effective electronic behavior has been observed.\cite{li02} 

\end{appendix}

\begin{acknowledgments}
The DFG (SFB 463), the DAAD (H.R.) and the NSF (DMR-0114818) are gratefully
acknowledged for financial support. We thank A. Ignatov for providing us with 
Fig. 11(b) and S. Savrasov, A. Ignatov, I. Mazin,
W. Pickett and T. Mishonov for discussions.


\end{acknowledgments}

\end{document}